%% file: MOR-arxiv.tex
\documentclass[10pt, letterpaper]{IEEEtran}
\input{preambleT.tex}
\usepackage[dvipsnames]{xcolor}

\title{Model Order Reduction for Water Quality Dynamics}
\author{Shen Wan$\text{g}^\dagger$, Ahmad F. Tah$\text{a}^{\ast}$, Ankush Chakrabart$\text{y}^\ddagger$, Lina Sel$\text{a}^{\spadesuit}$, and Ahmed Abokif$\text{a}^{\diamond}$
\thanks{$^\dagger$School of Cyberspace Security, Beijing University of Posts and Telcommunications. Email: shen.wang@bupt.edu.cn}
\thanks{$^\ast$Department of Civil and Environmental Engineering, Vanderbilt University, Nashville, TN, USA. Email: ahmad.taha@vanderbilt.edu}
\thanks{$^\ddagger$Research Scientist, Mitsubishi Electric Research Laboratories, Cambridge,  MA, USA. Email: achakrabarty@ieee.org}
\thanks{$^{\spadesuit}$Department of Civil, Architecture, and Environmental Engineering, The University of Texas at Austin. Email: linasela@utexas.edu}
\thanks{$^{\diamond}$Department of Civil, Materials, and Environmental Engineering, The University of Illinois Chicago. Email: abokifa@uic.edu}
\thanks{$^\ast$Corresponding author. }
\thanks{This work is partially supported by National Science Foundation under grants 1728629, 2015603, 2015671, 2151392, and 2015658.}}

\makeatletter
\def\endthebibliography{%
	\def\@noitemerr{\@latex@warning{Empty `thebibliography' environment}}%
	\endlist
}
\makeatother

\begin{document}
\maketitle

\setlength{\abovedisplayskip}{3.5pt}
\setlength{\belowdisplayskip}{3.5pt}
\setlength{\abovedisplayshortskip}{3.1pt}
\setlength{\belowdisplayshortskip}{3.1pt}

\newdimen\origiwspc%
\newdimen\origiwstr%
\origiwspc=\fontdimen2\font
\origiwstr=\fontdimen3\font

\fontdimen2\font=0.64ex

\begin{abstract}
A state-space representation of water quality dynamics describing disinfectant (e.g., chlorine) transport dynamics in drinking water distribution networks has been recently proposed. Such representation is a byproduct of space- and time-discretization of the PDE modeling transport dynamics. This results in a large state-space dimension even for small networks with tens of nodes. 
Although such a state-space model provides a model-driven approach to predict water quality dynamics, incorporating it into model-based control algorithms or state estimators for large networks is challenging and at times intractable.   To that end, this paper investigates model order reduction (MOR) methods for water quality dynamics with the objective of performing post-reduction feedback control. The presented investigation focuses on reducing state-dimension by orders of magnitude, the stability of the MOR methods, and the application of these methods to model predictive control.   
\end{abstract}



\section{Introduction and Paper Contributions}~\label{sec:literature}
\IEEEPARstart{W}{}ater quality models can be expressed in the form of a state-space representation~\cite{wang2020effective} representing the numerical solution of the PDEs describing the spatiotemporal evolution of disinfectant concentration (e.g., chlorine) in various network elements in drinking water distribution networks (WDN). Unfortunately, even for small-to-midsize networks, the dimension (or order) of the state-space models can reach $10^4$ or $10^6$, due to ensuring high fidelity in model predictions. 

This is due to discretization methods used for solving the partial differential equations (PDE) that describe the chlorine transport in pipes. Very high resolution in discretization yields improved accuracy in the control-oriented model but results in high state-space dimension, leading to heavy computational burden during simulation and difficulty for use in controller design~\cite{boulos2006comprehensive}. This implies that high-accuracy models, though effective for predicting system dynamics, are not amenable to controller or estimator design for large-scale networks---especially in the presence of state or input constraints. To that end, model order reduction (MOR) is necessary to derive a compact model for fast simulation and efficient synthesis of controllers and state estimators.

A \textit{full-order model} is referred to as an exact or near-exact model to describe water quality dynamics and usually is high in dimension. The motivation of MOR is to reduce the {full-order model}  to a \textit{reduced-order model} that has a much smaller number of states or order without significantly decreasing model accuracy while maintaining input-output relationships and retaining certain properties of the system such as controllability and observability. To this end, the high-dimensional state vector from the full-order model is projected onto a lower-dimensional subspace within which the model accuracy and certain properties are not significantly compromised. The MOR method has been widely applied in computational fluid dynamics~\cite{rowley2005model,Baur2014,Willcox2002}, and other fields such as computational electromagnetics, as well as in micro- and nano-electro-mechanical systems design~\cite{Montier2017}. However, reduced-order models have never been explored for modeling water quality dynamics---a gap that is filled in this paper.

Before delving into the details and literature of MOR algorithms, we note that the objective of this paper is \textit{not} to perform MOR for water quality (WQ) dynamics for the sake of observing state trajectories or simulations, seeing that efficient solvers have been developed for that purpose. The end-goal of  WQ MOR as we envision it in this paper is for it to be useful for post-MOR feedback control, among other applications.

\subsection{Literature review of MOR algorithms}
Various MOR algorithms have been developed in the rich literature of dynamic systems. The majority of these algorithms can be divided into two categories: singular value decomposition (SVD) based methods~\cite{adamjan1971analytic,Glover1984,moore1981principal,sirovich1987turbulence,Willcox2002,rowley2005model} and Krylov subspace methods or moment matching based methods~\cite{grimme1997krylov,antoulas1990solution,beattie2008interpolation,Bai2002,gallivan2006model}. There are also some studies combining SVD and Krylov methods~\cite{gugercin2008iterative}. In general, methods based on Krylov subspaces do not preserve important properties of the original system such as stability and passivity~\cite{Baur2014}. Hence, we are interested in SVD-based methods in which several methods can guarantee stability or controller designed related properties. Moreover, SVD-based approaches can be further divided into subcategories, for example, the methods using Hankel operators concepts~\cite{adamjan1971analytic,Glover1984}, the balanced realization theory~\cite{moore1981principal,lall2003error,rowley2005model}, and proper orthogonal decomposition (POD) based methods~\cite{lumley2007stochastic,sirovich1987turbulence,Willcox2002,rowley2005model,astrid2008missing}. This paper focuses on seeking proper SVD-based model reduction methods for the water quality dynamics represented linear discrete-time systems with a state-space presentation. 

Moore~\cite{moore1981principal}  proposes the balanced truncation (BT) method which considers or balances controllability and observability Gramians. Unfortunately, the BT method becomes computationally impractical for large-scale systems  (e.g., $10^4$ states or more). The  POD method proposed by Sirovich~\cite{sirovich1987turbulence} is tractable, but not as accurate as of the BT. To compute balancing transformations for high-order systems, Willcox~\cite{Willcox2002} proposes a technique that combines the POD and concepts from balanced realization theory. However, when the number of outputs of a system is large, this method is impractical, and it has the risk of truncating states that are poorly observable yet very strongly controllable. As a result, the balanced POD (BPOD)~\cite{rowley2005model} is proposed to address the same issues, and it combines the advantages of BT and POD methods successfully. That is, BPOD balances both controllability and observability and is tractable for a large-scale system, which results in the computational cost similar to POD and the accuracy similar to BT. 

In the water systems research community, the reduced models are applied in simplifying networks in the context of hydraulics and the techniques adopted are different from the ones aforementioned in dynamic systems. That is, the reduced model achieves a high-fidelity representation for the complete network hydraulic, but greatly simplifies the computation.  For example, a skeletonization method~\cite{ulanicki1996simplification} removes nodes from a linearized network model through Gauss elimination, and reduces the network model while preserving the nonlinear characteristics of the original network model. Salomons~\cite{shamir2008optimal} obtain a reduced model by the genetic algorithm. 
To meet the computational efficiency requirements of real-time hydraulic state estimation, Preis et al.~\cite{preis2011efficient} introduce a reduced model using a water system-aggregation technique. 
A methodology and application of a conjunctive hydraulic and water quality model for water distribution systems aggregation are presented in study~\cite{perelman2008water}. In contrast, this paper focuses on a different research problem that reduces the order of water quality dynamics model, which has not been studied in the literature, with the end-goal of performing post-MOR model driven control of water quality dynamics.

\subsection{Paper contributions}
The major objective of the paper is to investigate the performance of water quality MOR algorithms, address their theoretical limitations, assess their computational performance, and evaluate their potential when applied within a model-based predictive control framework. The below list outlines in detail the contributions of the paper. 
\begin{enumerate}
	\item We present the first attempt to identify reduced-order models for water quality simulation with theoretical analysis: our case studies show that the reduced-order model produces nearly identical water quality simulations in comparison with the full-order models. The MOR algorithms are tested for different scales of networks in sense of accuracy, low computational burden, robustness to initial conditions, and potential to be obtained as a stable reduced-order model.
	
	\item Classic MOR algorithms cannot ensure the stability of a reduced-order model, despite the stability of the underlying water quality dynamics. Thus, we propose two methods that guarantee that the reduced-order model is stable. The first method performs a stabilization process based on the standard MOR procedure while trying to maintain accuracy, and the second method simply adjusts a parameter in the standard MOR (i.e., BPOD) procedure. Moreover, the parameter adjustment has a clear physical interpretation for water quality dynamics, which is the largest travel time-step from booster stations to sensors.
	\item The obtained stable reduced-order models are then utilized as predictive models in MPC for a WDN. Our case study shows that the application of MPC to the reduced-order model produces similar control effects compared with the application of MPC to the full-order model, while the computational burden is significantly reduced.
\end{enumerate}

The rest of the paper is organized as follows. Section~\ref{sec:WDNModel} briefly describes control-oriented water quality modeling, then presents the full-order model in state-space form. The principle of MOR for the specific water quality modeling is given at the beginning of Section~\ref{sec:MOR}, followed by the fundamental procedures of various MOR methods and the corresponding discussions in Section~\ref{sec:comparsion}. Section~\ref{sec:stability} presents two methods to obtain a stable reduced-order model by adjusting the standard MOR procedures.  
Section~\ref{sec:casestudy} presents case studies that corroborate our proposed approach. The paper also comes with a few  limitations which are all acknowledged, jointly with future research directions, in Section~\ref{sec:limitations}.

\noindent \textit{Paper's Notation.} \hspace{0.2cm} Boldface characters represent matrices and vectors: $a$ is a scalar, $\m a$ is a vector, and $\m A$ is a matrix. Matrix $\m I$ denotes an identity matrix, whereas $\m 0_{m \times n}$ denotes a zero matrix  with size $m$-by-$n$.
The notation $\mathbb{R}$ denotes the set of  real numbers, and the notations $\mathbb{R}^n$ and $\mathbb{R}^{m\times n}$ denote a column vector with $n$ elements and an $m$-by-$n$ matrix in $\mathbb{R}$. For any vector $\m x \in \mathbb{R}^{n}$,  $\m x^{\top}$ is its  transpose. For any two matrices $\m A$ and $\m B$ with same number of columns, the notation $\{\m A, \m B\}$ denotes $[\m A^\top \  \m B^\top]^\top$. The norm of $\m A$ is $\norm {\m A}_2$. The $i$-th singular value of $\m A$ is denoted by  $\sigma_i$; a vector $\m \sigma(\m A)$ stands for all the singular values of $\m A$; the $k$-largest singular values of $\m A$ is denoted by vector $ \m \sigma^\downarrow_{k}(\m A)$. The $i$-th eigenvalue of $\m A$ is denoted by  $\lambda_i$; $\m \sigma(\m A)$ stands for all the eigenvalues of $\m A$; the $k$-largest eigenvalues of $\m A$ is expressed by vector $\m \lambda^\downarrow_{k}(\m A)$.

\section{State-Space Water Quality Model}~\label{sec:WDNModel}
In this section, we present the necessary background for the state-space model of water quality dynamics.
\subsection{Water quality dynamics}
We model WDN by a directed graph $\mathcal{G} = (\mathcal{N},\mathcal{L})$.  The set $\mathcal{N}$ defines the nodes and is partitioned as $\mathcal{N} = \mathcal{J} \cup \mathcal{T} \cup \mathcal{R}$ where $\mathcal{J}$, $\mathcal{T}$, and $\mathcal{R}$ are collection of junctions, tanks, and reservoirs. Let $\mathcal{L} \subseteq \mathcal{N} \times \mathcal{N}$ be the set of links, and define the partition $\mathcal{L} = \mathcal{P} \cup \mathcal{M} \cup \mathcal{V}$, where $\mathcal{P}$, $\mathcal{M}$, and $\mathcal{V}$ represent the collection of pipes, pumps, and valves. 

\begin{figure}
	\centering
	\includegraphics[width=0.95\linewidth]{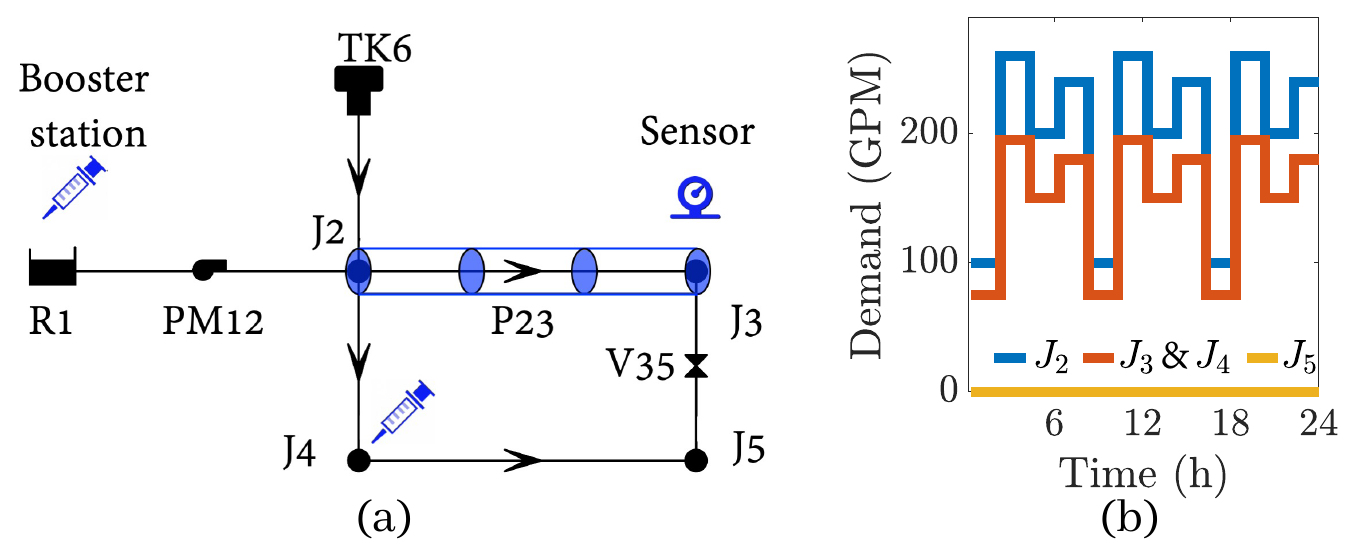}
	\caption{(a) exemplar topology of a WDN (multi-input and multi-output system with two boosters (inputs) and one sensor (output) to inject and detect chlorine), and $P23$ is split into three segments according to L-W scheme for illustration purpose; \textcolor{black}{(b) demand profiles for all junctions during 24 hours ($J3$ and $J4$ have the same demand pattern). In this figure, we show that (without loss of generality), demand values can occasionally be null for some junctions such as $J5$}. The WQ model should be able to take that as well as the inclusion of loops and booster stations at arbitrary junctions into account.}
\label{fig:network-demand}\vspace{-0.05cm}
\end{figure}

An example of a WDN graph is shown in Fig.~\ref{fig:network-demand}a, and a disinfectant (such as chlorine) is added into the network with a proper mass and concentration by booster stations installed at the reservoir source $R_1$ and the junction $J_4$ to disinfect the WDN. Each junction in a WDN is a type of node that connects two links. They may consume water, that is, they have water demands. During a specific period, demands from all junctions form a demand profile (see Fig.~\ref{fig:network-demand}b) which is assumed as a priori for water quality simulations in water research. Moreover, all components are designed to meet a certain threshold of water quantity and quality. For example, the pump $PM_{12}$ is designed to maintain the water pressure for all users (junctions).  The concentration sensors are installed at specific junctions for measurement purposes (see the sensor installed at $J_3$ in Fig.~\ref{fig:network-demand}a), in this way the water system operators can obtain the concentrations around the network.

Eventually, the water quality model can be used to represent the movement of all chemical and/or microbial species (contaminant, disinfectants, DBPs, metals, etc.) within a WDN as they traverse various components of the network. Specifically, the single-species interaction and dynamics of chlorine are considered in our paper, and it can be expressed using a state-space model. We first introduce the concepts and the corresponding methods that form the state-space model next.

In a WDN, the pipe model is represented by chlorine transport in differential pipe lengths by advection in addition to its decay due to reactions. Mathematically, for any pipe, the 1-D advection-reaction equation is given by a PDE. Although the PDE has no closed-form solution, the Lax-Wendroff scheme~\cite{lax1964difference} can be adopted to discretize and solve it numerically. For example, pipe $P_{23}$ in Fig.~\ref{fig:network-demand}a is split into three segments for illustrative purpose according to the scheme. In practice, the number of segments for a pipe (denoted by $s_{L}$) is decided by the flow velocity and the length of that pipe. After the discretization for all pipes, we define a vector $\m c^\mathrm{P}(t)$ to collect all segments in all pipes at time $t$. Similarly, the concentrations in pumps and valves at time $t$ are denoted by $\m c^\mathrm{M}(t)$ and $\m c^\mathrm{V}(t)$. The $\m c^\mathrm{L}(t) \triangleq \{\m c^\mathrm{P}(t), \m c^\mathrm{M}(t), \m c^\mathrm{V}(t) \} $ is a defined vector collecting all concentrations from links (i.e., pipes, pumps, and valves).  For all nodes, we define $\m c^\mathrm{N}(t) = \{\m c^\mathrm{J}(t), \m c^\mathrm{R}(t), \m c^\mathrm{T}(t)\} $ to collects the concentration from all junctions, reservoirs, and tanks.

\subsection{State-space representation}

In general, the number of junctions, reservoirs, tanks, pipes, pumps, and valves is denoted by $n_{\mathrm{J}}$, $n_{\mathrm{R}}$, $n_{\mathrm{TK}}$, $n_{\mathrm{P}}$, $n_{\mathrm{M}}$, and $n_{\mathrm{V}}$. Hence, the number of states in nodes and links are $n_\mathrm{N} = n_{\mathrm{J}}+n_{\mathrm{R}}+n_{\mathrm{TK}}$ and $n_\mathrm{L} = n_{\mathrm{P}} \times s_{L} +n_{\mathrm{M}}+n_{\mathrm{V}}$. Therefore,  $\m c^\mathrm{N}(t)  \in \mbb{R}^{n_\mathrm{N}}$ and $\m c^\mathrm{L}(t)  \in \mbb{R}^{n_\mathrm{L}}$.

Summarily,  a water quality state-vector $\m x$ defining the concentrations of the disinfectant (chlorine) in the network at time $t$ is denoted as $\m x(t) \triangleq  \{\m c^\mathrm{N}(t),\m c^\mathrm{L}(t) \} \in \mbb{R}^{n_x}$ and $n_x = n_\mathrm{N} + n_\mathrm{L}$. After applying the principle of conservation of mass,  we can rewrite all component models (i.e., water quality model) using a discrete-time state-space model of the form
	\begin{subequations}~\label{equ:fullmode-LTV}
		\begin{align}
		\m x(k+1) &= \m A(k)  \m x(k) + \m B(k)  \m u(k) ~\label{equ:fullmode_state-LTV}\\
		\m y(k) &= \m C \m x(k) + \m D   \m u(k),
		\end{align}
	\end{subequations} 
where $\m x(k) \in \mathbb{R}^{n_x}$ is the state vector, the vectors $\m u(k) \in \mathbb{R}^{n_u}$ and $\m y(k) \in \mathbb{R}^{n_y}$ are the system inputs and outputs (i.e., $n_u$ booster stations and $n_y$ concentration sensors), and $\m A(k) $ and $\m B(k) $
 are time-varying system matrices. Note that state matrix $\m A(k)$ stands for the pure system dynamic (without any boosters); the element in input matrix $\m B(k) $ represents locations of booster stations and corresponding injected flow rates that are both assumed as known. Output matrix $\m C$ is constant, and it indicates where the fixed WQ sensors are installed. Matrix $\m D$ describes the relationship or path directly from inputs to outputs, for example, some booster stations are equipped with a built-in concentration sensor (i.e., both the booster station and the sensor are at the same location) to calculate the injection of chlorine accurately. Usually, these two matrices $\m C$ and $\m D$ are fixed and time-invariant assuming fixed sensor placements. 

Note that these system matrices are usually updated from half hour to several hours according to the changing demands in demand profiles (e.g., demand changes every 2 hours in Fig.~\ref{fig:network-demand}b). That is, during a specific period in the demand profile (demands at all junctions do not change), the system matrices remain the same, and the discrete-time linear time-varying (DT-LTV) system~\eqref{equ:fullmode} degenerates into a nearly identical discrete-time linear time-invariant (DT-LTI) system with the exception that $\m A(k):=\m A$ and $\m B(k):=\m B$, i.e., these two state-space matrices become constant with a fixed hydraulic profile. 
	\begin{subequations}~\label{equ:fullmode}
	\begin{align}
	\m x(k+1) &= \m A(k)  \m x(k) + \m B(k)  \m u(k) ~\label{equ:fullmode_state}\\
	\m y(k) &= \m C \m x(k) + \m D   \m u(k),
	\end{align}
\end{subequations} 
Let us denote a generic $n_x$-th order DT-LTI system~\eqref{equ:fullmode} as the tuple $(\m A, \m B, \m C,\m D)$ for brevity.
In this way, the DT-LTV system~\eqref{equ:fullmode} comprises different tuples (DT-LTI systems). For example, DT-LTV models for 24 hours of the WDN in Fig.~\ref{fig:network-demand}a can be represented by 12 different DT-LTI systems since the demand changes each two hours.

It is worthwhile to note that \textit{(i)} the water quality model~\eqref{equ:fullmode} is available in our recent published work~\cite{wang2020effective}, and the model is stable; \textit{(ii)} after testing, each pipe is required to be split into at least 100 segments on average to guarantee the accuracy of discretizing PDEs. That is, the order $n_x$ can easily reach $10^4$ level for a mid-sized network with hundreds of pipes; \textit{(iii)} all system matrices in this model are sparse, and the sparsity rate could be from $99\%$ to $99.999\%$ or higher. Moreover, the challenging MOR problem for a DT-LTV system is equivalent to solving multiple relatively easy MOR problems for DT-LTI systems, and the detail is presented next.

\section{Model Order Reduction} ~\label{sec:MOR}
This section provides an overview of MOR principles for DT-LTI systems. For each specific MOR method, the corresponding procedure is summarized. 

\subsection{Principle of MOR for DT-LTI system}~\label{sec:principle}
The objective of MOR is to find a reduced-order model of order $n_r\ll n_x$, given by
	\begin{subequations}~\label{equ:reducedmode}
		\begin{align}
		\m x_r(k+1) &= \m A_r \m x_r(k) + \m B_r \m u(k)\\
		\hat{\m y}(k) &= \m C_r \m x_r(k) + \m D_r \m u(k),
		\end{align}
	\end{subequations}
where the new state vector $\m x_r \in \mathbb{R}^{n_r}$; the system input and output vectors $\m u $ and $\hat{\m y}$ retain the same dimension. From a simulation perspective, the estimated output $\hat{\m y}(k)$ from the reduced-order model is compared to the output from the full-order system $\m y(k)$ to test the performance of the MOR algorithms.

The general idea of the aforementioned reduction process is to find an invertible transformation matrix $\m T$ that maps state $\m x$ into another space state $\m z$. Specifically, $\m x = \m T \m z$, where the reduced-order model state $\m z \in \mathbb{R}^{n_x}$ and transformation matrix $\m T \in \mathbb{R}^{n_x \times n_x}$. Moreover, matrix $\m T$ is constructed based on some specific principles so that the elements in $\m z$ are ordered according to their \textit{importance}; more on that in Section~\ref{sec:fundamental}. After substituting $\m x = \m T \m z$ into~\eqref{equ:fullmode}, we obtain 
	\begin{subequations}~\label{equ:fullmode1}
		\begin{align}
		\m z(k+1) &=  \m A_z \m z(k) + \m B_z  \m u(k)\\
		\m y_z(k) &= \m C_z  \m z(k) + \m D_z \m u(k),
		\end{align}
	\end{subequations}
where system matrices $\m A_z = \m T^{-1} \m A \m T$, $\m B_z = \m T^{-1} \m B$, $\m C_z = \m C \m T$, and $\m D_z = \m D$.

However, the order of transformed system $(\m A_z, \m B_z, \m C_z,\m D_z)$ is still $n_x$. To obtain the reduced-order model~\eqref{equ:reducedmode}, only the first $n_r$ important elements in $\m z$ are kept and denoted by $\m x_r$. That is, $\m x$ can be replaced by $\m T_r \m x_r$ while the system properties remain the same, where $\m T_r \in \mathbb{R}^{n_x \times n_r}$ is the first $n_r$ columns of $\m T$. 
Finally, Equation~\eqref{equ:fullmode1} turns into reduced-order model~\eqref{equ:reducedmode} where 
	\begin{align}~\label{equ:transformation}
	\m A_r = \m S_r \m A \m T_r, \m B_r = \m S_r \m B, \m C_r = \m C \m T_r ,  \m D_r = \m D. 
	\end{align}
Note that $\m S_r$ in~\eqref{equ:transformation} is neither the inverse or pseudo inverse of $\m T_r$, it is simply the first $n_r$ row of $\m T^{-1}$.

In short, MOR is achieved by defining a certain subspace within the original space and transforming the system dynamics. A question that remains unanswered is which reduced-order models are most beneficial for achieving our goal: to reduce the difficulty of designing control/estimation algorithms such as Kalman filter, model predictive control (MPC), or linear–quadratic regulator (LQR) for water quality regulation. Since these control algorithms are related to the concepts of controllability and observability, one could argue that furthering our understanding of the controllable and observable subspaces could prove most useful for controller design. 

Controllability describes the ability of an external input $\m u$ to drive system state $\m x(k)$ from any initial state $\m x(0)$ to any other final state $\m x(k_f)$ in a finite time interval $[0,k_f]$~\cite{chen2013linear}. A quantitative metric for controllability in a DT-LTI system is the \textit{controllability Gramian} defined as 
	\begin{align}\label{equ:controllabilityGramian}
	\m W_C= \textstyle \sum_{m = 0}^{\infty} \m A^{m} \m B \m B^\top (\m A^\top)^{m},
	\end{align}
which is the solution of Lyapunov equation $\m W_C- \m A \m W_C\m A^\top = \m B \m B^\top$.
Similarly, observability describes the ability of reconstructing the initial unknown state $\m x(0)$ in finite $k_f$ steps from the knowledge of output $\m y(k)$~\cite{chen2013linear}. The corresponding metric for observability is the \textit{observability Gramian} defined as 
\begin{align}\label{equ:observabilityGramian}
	\m W_O= \textstyle \sum_{m = 0}^{\infty} (\m A^\top)^{m} \m C^\top \m C \m A^{m},
	\end{align} 
and it is the solution of Lyapunov equation $\m W_O- \m A^\top \m W_O\m A = \m C^\top \m C $.

The physical meaning of MOR in terms of using controllable and observable subspaces are that the uncontrollable (not affected by the input $\m u$) and unobservable (does not affect the output $\m y$) parts of the system without significantly affecting input-output relations are discarded. The methods considering controllability or observability Gramian are also referred to as \textit{Gramian based MOR}, and when both Gramians are considered in a method, it is then named as \textit{cross-Gramian based MOR}~\cite{Baur2014,Himpe2018}. Next, we present several specific MOR algorithms that consider the Gramians which are suitable for water quality application.

\subsection{Description of common MOR algorithms}~\label{sec:fundamental}
Herein, we describe three reduced-order modeling algorithms: balanced truncation, proper orthogonal decomposition, and balanced POD.
\subsubsection{Balanced Truncation}~\label{sec:BT}
Balanced trunction computes a specific $\m T$ that transforms the coordinate system $\m x$ into a new coordinate system $\m z$ in which the controllability and observability Gramians (denoted $\m W_{Cz}$ and $\m W_{Oz}$, respectively) are diagonal and identical~\cite{moore1981principal}. We denote this particular Gramian by $\m \Omega_z$. Mathematically, $\m \Omega_z = \m W_{Cz} = \m W_{Oz} $, which implies that $\m \Omega_z^2 = \m W_{Oz}  \m W_{Cz} $, where
	\begin{align*}
	\m W_{Cz} &= \textstyle \sum_{m = 0}^{\infty} \m A_z^{m} \m B_z \m B_z^\top (\m A_z^\top)^{m}  = \m T^{-1} \m W_C {\m T^{-1}}^\top, \notag 
	\end{align*}
and $\m W_{Oz}  =  \m T^\top\m W_O\m T$.

Note that $\m W_C$ and $\m W_O$ are the Gramians of the full-order model~\eqref{equ:fullmode} before transformation. Hence, 
	\begin{align*}
\m \Omega_z^2 = \m W_{Oz}  \m W_{Cz}  = \m T^{-1} \m W_O \m W_C\m T \Leftrightarrow \m W_O \m W_C\m T = \m T \m \Omega_z^2,
	\end{align*}
which indicates the transformation $\m T$ can be obtained by finding the eigenvector matrix of $\m W_O \m W_C$ (also known as \textit{cross Gramian matrix} $\m W_X$~\cite{Himpe2020}), and the $\m \Omega_z^2$ is just the corresponding diagonal eigenvalue matrix. The final $\m T_r$ can be obtained simply by selecting the first $n_r$ columns of $\m T$. 

Furthermore, the diagonal element $\sigma_i$, $i = {1,\ldots, n_x}$ in $\m \Omega_z$ measuring controllability and observability simultaneously are the \textit{Hankel singular values} (HSVs)~\cite{lall1999empirical} of $\m W_X$. Each single HSV $\sigma_i = \sqrt{\lambda_i}$ provides a measure of ``energy" for each state in a system. The model order reduction procedure provided by BT can also be viewed as retaining the $n_r-$largest HSVs (i.e., $\m \sigma^\downarrow_{n_r}(\m W_X)$) or eigenvalues (i.e., $\m \lambda^\downarrow_{n_r}(\m W_X)$)  with higher energy and discarding the least important ones with lower energy. The $n_r$ can be either specified as a fixed number directly or solved by specified energy level defined by  $ \mathrm{energy} = {\sum_{j = 1}^{n_r} \sigma_j}/{\sum_{i = 1}^{n_x} \sigma_i}$. The simplified procedure of BT approach is summarized as Procedure~\ref{proc:BT}.
\begin{procedure}[h]
	\small \DontPrintSemicolon
	Form  cross Gramian matrix $\m W_X = \m W_O \m W_C$ after solving  two Lyapunov equations\;
	Find the eigenvector matrix $\m T$ of $\m W_X$\;
	Specify $n_r$ as needed  or obtain $n_r$ via setting an energy level\;
	Extract $\m T_r$ from the first $n_r$ columns of $\m T$\;
	Extract $\m S_r$ from the first $n_r$ rows of $\m T^{-1}$\;
	Obtain reduced-order model~\eqref{equ:reducedmode} by~\eqref{equ:transformation}\;
	\caption{Classical BT()}
	\label{proc:BT}
\end{procedure}

This BT method considers and balances the controllability and observability simultaneously with high accuracy. Besides that, it preserves stability, which is a necessary property for the reduced-order system. However, it requires the computation of $\m W_O$ and  $\m W_C$ or equally solving two Lyapunov equations and eigenvalue decomposition that are \textcolor{black}{impractical}  for a large-scale network.

\subsubsection{Proper Orthogonal Decomposition}~\label{sec:POD} 
The POD algorithm~\cite{lumley2007stochastic,sirovich1987turbulence}, which is also known as principal component analysis or the Karhunen-Lo\`eve expansion, focuses on seeking a transformation $\m T_r$ such that a set of given data $\m x(t) \in \mathbb{R}^{n_x}$, with time $t \in [0,m]$ is projected into the subspace of fixed dimension $n_r$  as $\m x_r = \m T_r \m x \in \mathbb{R}^{n_r}$ while minimizing the total error $\sum_{t = 0}^{m-1} \norm {\m x(t)- \m x_r(t)}_2$. To solve this problem, the data is assembled into an $ n_x \times m$ matrix 
\begin{align}\label{equ:Snapshot_X}
\m X_m  = \begin{bmatrix}
\m x(0)  & \m x(1)   & \ldots & \m x(m-1)  
\end{bmatrix}.
\end{align} 
We define this matrix as the \textit{snapshot} matrix, where $m$ is the length of the snapshot (i.e., how many data samples are collected). Each vector in this matrix defines the state at a different time-instant (or sampling time). Define matrix $\m W_{Cm}$ as $\m X_m \m X_m^\top \in \mathbb{R}^{n_x \times n_x}$. Matrix $\m W_{Cm}$ is controllability Gramian-like matrix containing the first $m$ items when calculating $\m W_C$, and it is an approximation of $\m W_C$. We refer to $\m W_{Cm}$ as the \textit{$m$-step controllability Gramian}.  The optimal transformation $\m T_r$ in POD method can be obtained by eigenvalue decomposition $\m W_{Cm} \m T_r = \m T_r \m \Lambda $. Each eigenvector or the column vector in $\m T_r$ is named as \textit{POD mode}~\cite{rowley2005model}, and the eigenvalues of $\m W_{Cm}$ are the diagonal elements of $\m \Lambda$.

Note that a snapshot is given by~\eqref{equ:Snapshot_X} which, if $\m A$ is sparse, can be easily obtained even with $n_x \geq 10^6$ since the computational task involves sparse matrix multiplication. However, finding the eigenvectors of an $n_x \times n_x$ matrix $\m W_{Cm}$ is still challenging. 
Sirovich~\cite{sirovich1987turbulence} solves this problem by defining $\widetilde{\m W}_{Cm} = \m X_m^\top \m X_m $ which is in $\mathbb{R}^{m \times m}$ instead of $\mathbb{R}^{n_x \times n_x}$ with $m \ll n_x$. The matrix form of eigenvector and eigenvalue of this new $\widetilde{\m W}_{Cm}$ are denoted as $\m U$ and $\m \Lambda$.  That is, the $n_x \times n_x$ eigenvalue problem is solved thorough constructing an $m \times m$ eigenvalue problem. Finally, the transformation matrices $\m T_r $ and  $\m S_r $ can be obtained, and the detail is in Procedure~\ref{proc:POD}. Note that when $n_x \ll m$, the above step is not necessary, and computing $\m W_{Cm}$ directly is easier and preferred.
\begin{procedure}
	\small \DontPrintSemicolon
	Construct snapshot $\m X_m$ using \eqref{equ:Snapshot_X}\;
	Perform eigenvalue decomposition for $\widetilde{\m W}_{Cm} = \m X_m^\top \m X_m $ by finding $\m U$ and $\m \Lambda$ via $\widetilde{\m W}_{Cm} \m U = \m U  \m \Lambda$\;
	Specify $n_r$ as needed  or obtain $n_r$ via setting an energy level\;
	Extract transformation $\m T_r$ from the first $n_r$ column of $\m X_m \m U \m \Lambda^{-\frac{1}{2}}$\;
	Extract transformation $\m S_r$ from the first $n_r$ rows of $\m \Lambda^{-\frac{1}{2}}  \m U \m X_m^\top$\;
	Obtain reduced-order model~\eqref{equ:reducedmode} by~\eqref{equ:transformation}\;
	\caption{Classical POD()}
	\label{proc:POD}
\end{procedure}

Since the eigenvalues of $\m A \m B$ are the same as those of $\m B \m A $ (see~\cite[Theorem 1.3.22]{horn2012matrix}), i.e., $\m \lambda (\m A \m B) = \m \lambda (\m B \m A) $, the $\widetilde{\m W}_{Cm}$ shares the same eigenvalues with {$m$-step controllability Gramian} $\m W_{Cm}$. That is, POD, not as the BT method that exploits the controllability Gramian (summations of infinity items), uses only the first $m$ items of the controllability Gramian. This makes POD tractable for even large-scale systems but less accurate compared to BT.  Thus, POD only considers controllability Gramian and selects the $n_r$ modes with high-energy states (i.e., $\m \lambda^\downarrow_{n_r}(\m W_{Cm})$), and discards the low-energy states.

\subsubsection{Balanced Proper Orthogonal Decomposition}~\label{sec:BPOD} 
BPOD aims to combine the advantages of BT and POD together, that is, to obtain an approximation to BT that is {computationally} tractable for large-scale systems~\cite{rowley2005model}. Different from the snapshot $\m X_m$~\eqref{equ:Snapshot_X} defined in POD method, BPOD defines  $\m X_m$ as
	\begin{align}\label{equ:Snapshot_X_BPOD}
	\begingroup 
	\setlength\arraycolsep{1pt}
	\hspace{-0.5em} \m X_m  = \begin{bmatrix}
	\m x_1(0) & \ldots & \m x_1(m\hspace{-2pt}-\hspace{-2pt}1) & \ldots  &  \m x_{n_u}(0) & \ldots & \m x_{n_u}(m\hspace{-2pt}-\hspace{-2pt}1)  
	\end{bmatrix}
	\endgroup,
	\end{align} 
where the $\m x_i(m)$ is the impulse response (when $\m u_i(k) = \delta(k)$ is an impulse signal). That is,  $\m x_i(m) = \m A^{m} \m b_i$ and $\m b_i$ is the $i$-th column of $\m B$.

Moreover, BPOD denotes  $\m Y_m$ as the snapshot of the dual system of system $(\m A, \m B, \m C, \m D)$ or equally~\eqref{equ:fullmode}. That is, $\m Y_m$ is the snapshot of system $(\m A^\top, \m C^\top, \m B^\top,  \m D^\top)$ with system state denoted as $\m z$, and $\m Y_m$ can be expressed  as
	\begin{equation}\label{equ:Snapshot_Y_BPOD}
	\begingroup 
	\setlength\arraycolsep{1pt}
	\hspace{-0.5em}  \m Y_m  = \begin{bmatrix}
	\m z_1(0) & \ldots & \m z_1(m\hspace{-2pt}-\hspace{-2pt}1) & \ldots  &  \m z_{n_y}(0) & \ldots & \m z_{n_y}(m\hspace{-2pt}-\hspace{-2pt}1)  
	\end{bmatrix}
	\endgroup,
	\end{equation} 
where the $\m z_i(m)$ is the impulse response, that is,  $\m z_i(m) = {(\m A^\top)}^{m} \m c_i^\top$ and $\m c_i^\top$ is the $i$-th column of $\m C^\top$.

Then, the balancing modes are computed by forming the singular value decomposition of the \textit{block Hankel matrix} $ \m H_m$:

\begingroup
\setlength\arraycolsep{1pt}
	\begin{align}~\label{equ:svd}
	\m H_m = \m Y_m^\top \m X_m = \m U \m \Sigma \m V^\top = \begin{bmatrix}
	\m U_r & \m 0
	\end{bmatrix} \begin{bmatrix}
	\m \Sigma_r & \m 0\\
	\m 0 & \m 0
	\end{bmatrix} \begin{bmatrix}
	\m V_r^\top \\ \m 0 
	\end{bmatrix},
	\end{align} 
\endgroup

where the diagonal element in $\m \Sigma_r \in \mathbb{R}^{n_r \times n_r}$ collects the largest $n_r$ singular values of $\m H_m$, and $\m U_r$ and $\m V_r^\top$ are the corresponding left- and right-singular vectors. Hence, the final transformation matrices are $\m T_r = \m X_m \m V_r \m \Sigma_r^{-\frac{1}{2}}$ and $\m S_r = \m \Sigma_r^{-\frac{1}{2}} \m U_r^\top \m Y_m^\top$; see Procedure~\ref{proc:BPOD} for details.
\begin{procedure}[h]
	\small \DontPrintSemicolon
	Form snapshots $\m X_m$~\eqref{equ:Snapshot_X_BPOD} and $\m Y_m$~\eqref{equ:Snapshot_Y_BPOD} of BPOD\;
	Perform singular value decomposition of  $\m H_m = \m Y_m^\top \m X_m $~\eqref{equ:svd}, and obtain $\m U_r$, $\m \Sigma_r$, $\m V_r$\;
	Specify $n_r$ as needed  or obtain $n_r$ via setting an energy level\;
	Calculate $\m T_r = \m X_m \m V_r \m \Sigma_r^{-\frac{1}{2}}$ and $\m S_r = \m \Sigma_r^{-\frac{1}{2}} \m U_r^\top \m Y_m^\top$\;
	Obtain reduced-order model~\eqref{equ:reducedmode} by~\eqref{equ:transformation}\;
	\caption{Classical BPOD()}
	\label{proc:BPOD}
\end{procedure}

\begin{table*}[t]
	\centering
	\small
	\caption{Comparisons among three classical MOR methods.}
	\setlength{\tabcolsep}{4pt} 
	\renewcommand{\arraystretch}{1.3} 
	\label{tab:complexity}
	\begin{tabular}{c|c|c|c|c|c|c|c|c}
		\hline
		\textit{} & \textit{\makecell{Matrix \\ involved}} & \textit{\makecell{Dimension}} & \textit{\makecell{Decomposi- \\ tion method$^\dagger$}} & \textit{\makecell{Mode \\retained}} & \textit{\makecell{ Complexity \\(Computational load) }}  & \textit{\makecell{Large-scale \\tractability}} &   \textit{\makecell{Relative \\ accuracy$^\diamond$}}  & \textit{\makecell{Stability$^*$\\ preserving}} \\ \hline
		\textit{BT} & $\m W_X$ & $\mathbb{R}^{n_x \times n_x}$ & \textit{ED} & $\m \lambda^\downarrow_{n_r}(\m W_X)$ & $\mathcal{O}(n_x^3)$ (High) & \textit{No} & \textit{High} & \textit{Yes} \\ \hline
		\textit{POD$^\ddagger$} &  $\widetilde{\m W}_{Cm}$ & $\mathbb{R}^{m \times m}$ & \textit{ED} & $\m \lambda^\downarrow_{n_r}(\widetilde{\m W}_{Cm})$  & $\mathcal{O}(m^3)$ (Low) & \textit{Yes} & \textit{Low} & \textit{No} \\ \hline
		\textit{BPOD} & $\m H_m$ & $\mathbb{R}^{m n_y \times m n_u}$   & \textit{SVD} & $\m \sigma^\downarrow_{n_r}(\m H_m)$  & $\mathcal{O}(m^2 n_y n_u n_r)$ (Low) & \textit{Yes} & \textit{Medium} & \textit{{Depends}$^\spadesuit$} \\ \hline \hline
		\multicolumn{9}{l}{\makecell{$^\dagger$Eigenvalue decomposition (ED); singular value decomposition (SVD); $^\ddagger$use $\widetilde{\m W}_{Cm}$ when $m \ll n_x$, otherwise, use $\m W_{Cm}$ instead.\\ $^\diamond$Under the same number of modes; $^*$asymptotically stable;$^\spadesuit$Rowley~\cite{rowley2005model} conjectures that BPOD can preserve stability via theoretical analysis \\while other  studies~\cite{Montier2017,amsallem2012stabilization} denied this assertion based on test results.}}
	\end{tabular}
\end{table*}      

From the above procedures, it is clear that BPOD considers both controllability and observability Gramians like BT method, successfully avoids solving the two Lyapunov equations, forming cross Gramian matrix $\m W_X$, and finding the eigenvalues of $\m W_X$ to form a traceable algorithm simply by adopting the concept of the snapshot.

\section{Thoeretical Comparisons of MOR Algorithms for Water Quality Dynamics}~\label{sec:comparsion}
The advantages and disadvantages from the aspects of tractability for large-scale systems, relative accuracy, and stability preserving properties are summarized in this section. Subsequently, we present a comparison of these MOR algorithms to determine which properties of these algorithms fit best for our application--- the water quality model.

First, the complexity of solving Lyapunov equations~\cite{li2002low}, eigenvalue decomposition~\cite{pan1999complexity}, and SVD~\cite{feng2018faster} are  roughly estimated by $\mathcal{O}(n^3)$ for an $n \times n$ matrix. Although all BT, POD, and BPOD procedures include at least one of the above time-consuming computation, the matrix size appeared in these procedures and the number of executions of the procedures are different; see Tab.~\ref{tab:complexity}. For example, BT solves two Lyapunov equations ($n_x \times n_x$) and performs eigenvalue decomposition for $\m W_X $ once, that is, it performs algorithm with $\mathcal{O}(m^3)$ three times; POD performs eigenvalue decomposition for $\widetilde{\m W}_{Cm} $, the complexity is roughly $\mathcal{O}(m^3)$; BPOD performs truncated SVD (rank $n_r$; see~\eqref{equ:svd}) for $\m H_m$, the complexity is roughly $\mathcal{O}(m^2 n_y n_u n_r)$. Due to $m \ll n_x$, it is clear that BT is the slowest algorithm. The computational load of   POD and BPOD algorithms depends on the value of  $m$ and  $n_y n_u n_r$. When these values are of the same order of magnitude, the complexity of  POD and BPOD are similar and also tractable for large-scale systems; see Tab.~\ref{tab:complexity} for the final computational load.

Second, a useful property of the BT method is that the error bounds provided are close to the lower bound achieved by any reduced-order model, and the upper bound of the error is guaranteed less than $2 \sum_{i = n_r+1}^{n_x} \sigma_i$, where $\sigma_i$ is the $i$-th HSV of $\m W_X $ in Section~\ref{sec:BT}~\cite{rowley2005model}. Furthermore, BT considers $\m W_X$ that balances controllability and observability. Hence, the relative error in terms of preserving input-output relations of the BT method is theoretically the best among these three methods. As for POD, it only takes advantage of $m$-step controllability Gramian $\m W_{Cm}$, and has the risk of discarding the states that are highly observable but less controllable. This defect leads to the inaccuracy of POD or even the failure to reduce system order.

In fact, BPOD provides a good approximation of BT since they retain the same states/modes when performing MOR. The corresponding  $n_r$-largest eigenvalues (singular values) of all modes/states retained from three procedures are summarized in Tab.~\ref{tab:complexity}. That is, the states retained are those with  $\m \lambda^\downarrow_{n_r}(\m W_X)$ ($\m \sigma^\downarrow_{n_r}(\m H_m)$) for BT (BPOD). Suppose $\lambda_i$ ($\sigma_i$) is the $i$-th element of $\m \lambda^\downarrow_{n_r}(\m W_X)$ ($\m \sigma^\downarrow_{n_r}(\m H_m)$), and  $\lambda_i \approxeq \sigma_i^2$ always holds true when $m$ is large enough. 
{This implies that} BPOD, like BT, uses the same information (i.e., cross Gramians $\m W_X$) to obtain the transformation $\m T_r$ to reduce system order. The difference {lies} in the way {each algorithm deals} with such information. For example, BT {requires} an eigenvalue decomposition {of} $\m W_X$ while BPOD {involves} singular value decomposition to $\m H_m$ embedded in $\m W_{X}$. From this perspective, the transformation $\m T$ obtained from the BPOD method is the balanced one that simultaneously diagonalizes $\m W_C$ and $\m W_O$. Hence, BPOD has the same ability (when $m$ is large enough) ensure the accuracy as BT does.

The BT method guarantees stability theoretically after MOR~\cite{moore1981principal}, while POD fails~\cite{Baur2014,prajna2003pod} due to the projection it uses. As for the BPOD method, different studies report various results. Rowley~\cite{rowley2005model} claims that the stability-preserving from BPOD is guaranteed for linear systems when snapshots are large enough. 
However, there is no guideline to find what is the proper size of snapshots. We suspect that this might be why the results from several studies~\cite{Montier2017,amsallem2012stabilization} do not support the stability-preserving property of BPOD. In other words,  stability-preserving of BPOD depends on the parameter (i.e., length of snapshots) set. We would present an approach to find the proper parameter that ensures stability in Section~\ref{sec:stability}.


In short, POD and/or BPOD methods are still preferred since they are tractable for a large-scale system, even if their stability-preserving properties are not sufficient for water quality dynamics MOR. Toward that goal, the stability-preserving model order reduction approaches for water quality modeling are explored and proposed next. 

\section{Stabilizing MOR Algorithms}~\label{sec:stability}
In this section, the brief literature of stability-preserving MOR is presented first followed by proposing two methods suitable for the stable and discrete-time system. 


The approaches developing stability-preserving reduced-order model involve \textit{a priori}  frameworks~\cite{Rowley2004,Barone2009,kalashnikova2010stability,kalashnikova2012stable,sirisup2004spectral}, and \textit{a posteriori} frameworks~\cite{wang2012proper,kalashnikova2014stabilization,Bond2008,benosman2017robust}.
{\textit{A priori} stability-preserving framework can be difficult to implement, and/or requires the solution of PDEs or special system structure. For example, the Galerkin projection (i.e. a congruence transform) needs sign-definite matrices in the full-order system. Conversely, \textit{a posteriori} methods minimally modify the reduced-order system matrix so that the accuracy is not significantly changed. The more detailed literature can be found in~\cite{kalashnikova2014stabilization}. Herein, we discuss stability-preservation with both these frameworks.}

In short, we propose a novel a posteriori method to stabilize the reduced-order model after performing the POD or BPOD procedure. Comparing to existing methods such as~\cite{Bond2008}, the proposed a posteriori method adjusts the final transformation $\m A_r$ directly instead of optimizing/adjusting the projection $\m U_r$ slightly while fixing the projection $\m V_r^\top$ in~\eqref{equ:svd}. Furthermore, published studies focus more on continuous systems while the proposed one is for discrete systems and uses a special trick to relax the non-convex optimization problem.

Moreover, a priori stabilization is proposed specifically for the BPOD procedure. To the best of our knowledge, this is the first a priori stabilization method for BPOD. Comparing to existing studies such as~\cite{Rowley2004} focusing on finding a special and complex projection for POD, the proposed method is simple and intuitive since it only requires modifying one parameter in BPOD. 


\subsection{{A posteriori} stabilization for POD-based approach}~\label{sec:posterior} 
After performing the POD procedure, the stability of the reduced-order model cannot be guaranteed. That is, $\m A_r$ is unstable. Furthermore, the stability of BPOD depends on the length of snapshot $m$. When $m$ is not proper,  $\m A_r$ from BPOD tends to be unstable. The motivation behind this posterior method is to adjust $\m A_r$ by $\Delta \m A_r$ such that the adjusted system is stable while ensuring that the norm $\norm{\Delta \m A_r}_2$ is as small as possible to maintain the accuracy of the full-order system. 

From Lyapunov stability conditions~\cite{lyapunov1992general}, we know that a DT-LTI system $(\m A, \m B, \m C, \m D )$ is stable when there exist a positive definite matrix $\m P$ such that $\m A^\top \m P \m A - \m P$ is negative definite. Thus, in our case the stabilization problem can be expressed as an optimization problem below.
	\begin{subequations}~\label{equ:stablization}
		\begin{align}
		\min_{\m P,\Delta \m A_r,\alpha} \quad &  \alpha \\
		\subjectto \quad &  \m P \succ 0, \alpha > 0\\
		& \norm{\Delta \m A_r}_2 \leq \alpha \label{equ:limit}\\
		&  (\m A_r + \Delta \m A_r)^\top \m P (\m A_r + \Delta \m A_r) - \m P \prec 0, \label{equ:neg_sd}
		\end{align}
	\end{subequations}
where $\m P$, $\Delta \m A_r$ in $\mathbb{R}^{n_r \times n_r}$ and $\alpha \in \mathbb{R}$ are the optimization variables. The constraint $\m P \succ 0$ entails requiring matrix $\m P$ to be symmetric positive definite matrix, i.e., has positive eigenvalues. The constraint $ \norm{\Delta \m A_r}_2 \leq \alpha $ requires matrix variable $ {\Delta \m A_r}_2 $ to have 2-norm smaller or equal to another scalar variable $\alpha$. This problem~\eqref{equ:stablization} is nonconvex, and to deal with the nonconvexity we can take the following steps to transform it into an convex semidefinite programming (SDP). 

Since $\m P \succ 0$ is invertible, let $\m P = \m P \m P^{-1}  \m P $, and taking Schur complements, we have
	\begin{align}
	~\eqref{equ:neg_sd}  & \Leftrightarrow     (\m A_r + \Delta \m A_r)^\top \m P \m P^{-1}  \m P  (\m A_r + \Delta \m A_r) - \m P \prec 0 \notag \\
	& \Leftrightarrow  \begin{bmatrix}
	-\m P & (\m P \m A_r + \Delta \m Y)^\top \\
	\m P \m A_r + \Delta \m Y & -\m P
	\end{bmatrix} \succ 0, ~\label{equ:LMI1}
	\end{align} 
where $\Delta \m Y =  \m P \Delta \m A_r $. Furthermore, the Cauchy-Schwarz inequality yields 
	\begin{align*}\norm{\Delta \m Y}_2 = \norm{\m P \Delta \m A_r}_2 \leq \norm{\m P}_2 \norm{\Delta \m A_r}_2 .\end{align*}
Invoking~\eqref{equ:limit} yields $\norm{\Delta \m Y}_2\le \alpha \norm{\m P}_2$. Taking Schur complements yields	
	\begin{align*}
	\eqref{equ:limit}  & \Leftrightarrow   \norm{\Delta \m Y}_2 \leq \alpha_y     \Leftrightarrow  \begin{bmatrix}
	\m I & *  \\
	\Delta \m Y &  \alpha_y^2 \m I
	\end{bmatrix} \succeq 0,
	\end{align*}
where $\alpha_y = \alpha \norm{\m P}_2$. However, the $\alpha_y^2$  is still nonlinear and can be relaxed as
	\begin{equation}
	\begin{bmatrix}
	\m I & * \\
	\Delta \m Y &  \alpha_y \m I
	\end{bmatrix} \succeq 0, \;\; \alpha_y >0 ~\label{equ:LMI3}.
	\end{equation} 
Finally, a convex SDP stabilizing the reduced-order model is formed as
	\begin{subequations}~\label{equ:convex_stablization1}
		\begin{align}
		\min_{\m P,\Delta \m Y,\alpha_y} \quad &  \alpha_y \\
		\subjectto \quad &  \m P \succ 0,~\eqref{equ:LMI1},~\eqref{equ:LMI3}.
		\end{align}
	\end{subequations} 

Note that the dimension of optimization variables such as $\m P$ and $\Delta \m Y$ depends on the size of $\m A_r$ in reduced-order model~\eqref{equ:reducedmode} instead of the size of $\m A$ in full-order model~\eqref{equ:fullmode}. This indicates the formulation~\eqref{equ:convex_stablization1} can be applied to solve the stabilization problem even for a large-scale system, provided that $n_r$ is not large after reduction procedures.


\subsection{A priori stabilization for BPOD approach}~\label{sec:priori}
The a priori stabilization method introduced in this section is different from the posterior one since it directly modifies the length of snapshots $m$ in the standard BPOD  Procedure~\ref{proc:BPOD} while generating a stable reduced-order model.
%
Ideally, when parameter $m$ is set as $\infty$, the procedure then has all system information as the cross-Gramian does in the BT method, and consequently BPOD reaches its best performance. 
However, parameter $m$ cannot be too large (or equal to infinity) due to the prohibitive computational cost. 
Hence, selecting a lower bound for $m$ denoted by $\underline{m}$ is important.  To the best of our knowledge, this is the first time attempt to produce such a value or condition that ensures the stability of a DT system using BPOD method.  In particular, we propose two methods to find $\underline{m}$. The first method is generated from insights from transport dynamics in water networks. The second method is via a control theoretic approach. In the case studies section, we compare the performance of these two methods. 

\begin{figure}
	\centering
	\includegraphics[width=0.97\linewidth]{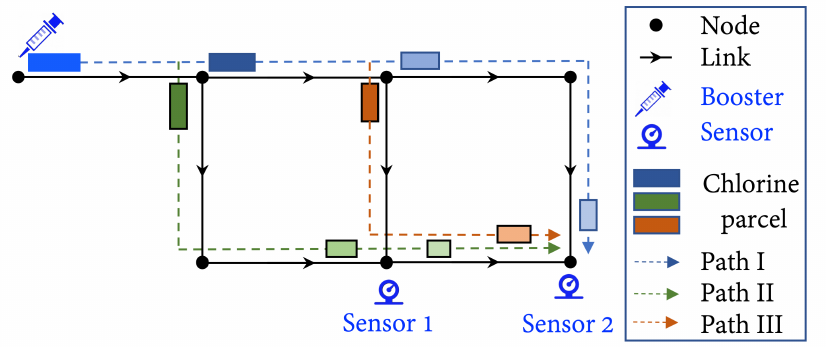}
	\caption{Chlorine travel paths from the input location (booster station) to different output locations (Sensors 1 and 2).}
	\label{fig:travel}
\end{figure}
\subsubsection{First method}
The minimum number of steps $\underline{m}$ is the equivalent time-step for a chlorine \textit{parcel} traveling from its input location to the furthest output location. Mathematically, it can be expressed as
	\begin{align}~\label{equ:travel}
	\underline{m} =  \left \lceil{
		\frac{T_{\mathrm{travel}}}{\Delta t}} \right \rceil  = \left \lceil{ {\sum} \frac{L_{ij}}{v_{ij} \Delta t} } \right \rceil , ij \in \mathcal{L_{\mathrm{path}}},
	\end{align}
where the notation $\left \lceil x \right \rceil$ denotes the ceiling of a real number $x$ (i.e., ${\displaystyle \lceil x\rceil =\min\{n\in \mathbb {Z} \mid n\geq x\}}$ where $\mathbb{Z}$ is the set of integers); $\Delta t$ is determined by the L-W scheme stability condition; the total travel time is denoted by $T_{\mathrm{travel}}$; $\mathcal{L}_\mathrm{path}$ represents the links in the travel path, and the corresponding length and velocity of each pipe $ij$ in that path are $L_{ij}$ and $v_{ij}$.


To illustrate the application of this method, we use the network comprised of two output sensors and one booster station controller; see Fig.~\ref{fig:travel}. After a chlorine parcel with a certain mass and concentration is injected at the booster station, it travels to the sensors with various velocities in different links. Note that both the mass and concentration of the chlorine parcel reduce gradually due to consumption by the nodes and decay of chlorine concentration. Furthermore, parcels (the rectangle in Fig.~\ref{fig:travel}) with different masses and concentrations (the areas and colors of the rectangles) arrive at the sensors at various times. The number of parcels depends on the number of paths from an input location (booster) to output one (sensor). For example, there are three paths in Fig.~\ref{fig:travel} and the furthest location from the input is Sensor 2, there are three travel times (three paths) for it from the booster, and the longest travel time should be used when calculating~\eqref{equ:travel}. 

\subsubsection{Second method}
From a control theoretic perspective, the lower bound $\underline{m}$ can also be estimated by the \textit{settling time} of water quality dynamic model~\eqref{equ:fullmode}. 
The settling time $T_s$ of a dynamic system can be roughly computed by its dominant pole $p$ which is equal to the eigenvalue of matrix $\m A$ with the largest magnitude. In particular, $T_s$ can be computed as 
	\begin{align*}
	T_s = \frac{-4 \Delta t}{\ln \abs{p}},
	\end{align*}
where $\Delta t$ is the sampling time of DT-LTI system~\eqref{equ:fullmode}; the dominant pole $p$ located within a unit disk indicates that $\ln \abs{p}$ is always negative; and $\abs{p}$ is the magnitude of the complex number $p$.  
Consequently, the lower bound of $m$ is
	\begin{align}~\label{equ:condition}
	\underline{m} =\left \lceil{  \frac{T_s}{\Delta t}} \right \rceil = \left \lceil{  \frac{-4 }{\ln \abs{p}}}\right \rceil.
	\end{align}

Note that the equivalent travel time-step in~\eqref{equ:travel} has a physical meaning for water quality dynamics which can be computed easily. Meanwhile, the second method is estimated by the empirical formula~\eqref{equ:condition}.  Section~\ref{sec:result-net1} investigates the performance of these two methods.

Finally, the overall procedure of general, stability-preserving MOR procedure is summarized in Algorithm~\ref{alg:generalMOR}. In this algorithm, we use BPOD but roughly the same approach can be applied to POD. 
	\begin{algorithm}[h]
		\small	\DontPrintSemicolon
		\KwIn{Full-order model $(\m A, \m B, \m C, \m D)$~\eqref{equ:fullmode} and control input $\m u(k)$}
		\KwOut{Stable reduced-order model $(\m A_r, \m B_r, \m C_r, \m D_r)$~\eqref{equ:reducedmode}, outputs $\m y(k)$ and $\hat{\m y}(k)$}
		\If{a posterior stabilization method}{
			Standard BPOD approach given in Procedure~\ref{proc:BPOD}\;
			Solve a convex SDP problem~\eqref{equ:convex_stablization1} to  stabilize the reduced-order model $(\m A_r, \m B_r, \m C_r, \m D_r)$\;
		}
		
		\If{a priori stabilization method}{
			Obtain parameter $m$ via~\eqref{equ:travel} or~\eqref{equ:condition}, and set length of snapshot as $m = \underline{m}$\;
			Execute BPOD procedure  to obtain reduced-order models $(\m A_r, \m B_r, \m C_r, \m D_r)$~\eqref{equ:reducedmode}\;
		}
		Generate $\m y(k)$ and $\hat{\m y}(k)$ from the full- and reduced-order models
		\caption{Stability-Preserving BPOD (SBPOD).}
		\label{alg:generalMOR}
	\end{algorithm}

\begin{table}[t]
	\caption{Basic information of three tested networks.}
	\small
	\centering
	\label{tab:info}
	\setlength\tabcolsep{3pt}
	\renewcommand{\arraystretch}{2}
	\begin{tabular}{c|c|c|c|c}
		\hline
		{\textit{Networks}} & {\textit{\makecell{\# of com-\\ponents$^*$}} } & {\textit{\makecell{Full order\\ $n_x$}}} & {\textit{\makecell{Booster station\\  locations ($n_u$)}}}  & {\textit{\makecell{Sensors\\ location ($n_y$)}}}   \\ \hline
		\textit{\textit{\makecell{three-node\\network}}} & \makecell{\{1,1,1,\\1,1,0\}}  & 154 & \makecell{J2 \\(1)} & \makecell{TK3 \\ (1)}    \\ \hline
		\textit{Net1} & \makecell{\{9,1,1,\\12,1,0\}}   & 1,293 &  \makecell{J10\\ (1)}  &  \makecell{J22, J23 \\ (2) }   \\ \hline
		\textit{Net3} & \makecell{\{92,2,3,\\117,2,0\}}  & 29,374 & \makecell{ J237, J247\\(2)}  & \makecell{J255, J241\\  J249 (3) }  \\ \hline \hline
		\multicolumn{5}{l}{\footnotesize{
				\makecell{$^*$Number of each component in WDN: \{$n_\mathrm{J}$, $n_\mathrm{R}$, $n_\mathrm{TK}$, $n_\mathrm{P}$, $n_\mathrm{M}$, $n_\mathrm{V}$\}.} }}
	\end{tabular}%
	\vspace{-0.4cm}
\end{table}

\section{Case Studies}~\label{sec:casestudy}
We present three examples (three-node, Net1, and Net3 networks~\cite{rossman2000epanet}) to illustrate the performance of three different MOR methods in terms of accuracy, computational load/time, and stability-preserving properties.  Then, we test the two stabilization methods proposed in Section~\ref{sec:stability}. Finally, we showcase the application of MOR algorithms when MPC is used to control water quality dynamics for the full-order and reduced-order systems. All codes, parameters and tested networks simulated via EPANET Matlab Toolkit~\cite{Eliades2016} are available on Github~\cite{wangMOR}. The testing environment is a Win 10 Precision 7920 Tower with Inter(R) Xeon(R) Gold 5218 CPU @2.3GHz and 64G memory.  

We note here that the objective of the first part of the case studies (where we test the performance of MOR algorithms and the proposed SBPOD) is to verify their performance in terms of their fidelity of predicting water quality dynamics. To do that, a variety of control-theoretic testing scenarios and conditions are implemented. Such testing scenarios are needed to verify the fidelity of the reduced order model, which is then used to perform feedback control and water quality regulation using model predictive control.  In particular, we test the different algorithms via different initial conditions of chlorine concentrations which does have an impact on how the methods perform. We also present results for step response, which refers to applying a constant control input $\m u(t)$ (injected chlorine mass from booster stations) to test the performance of various algorithms. The step response is widely used in dynamic system sciences, as many control inputs can be viewed as a sequence of changing constant signals resembling a staircase signals, comprised of varying magnitudes of control inputs. Finally, the different networks we test include different network topology and number and location of booster stations and water quality sensors.


\begin{figure}
	\centering
	\includegraphics[width=0.9\linewidth]{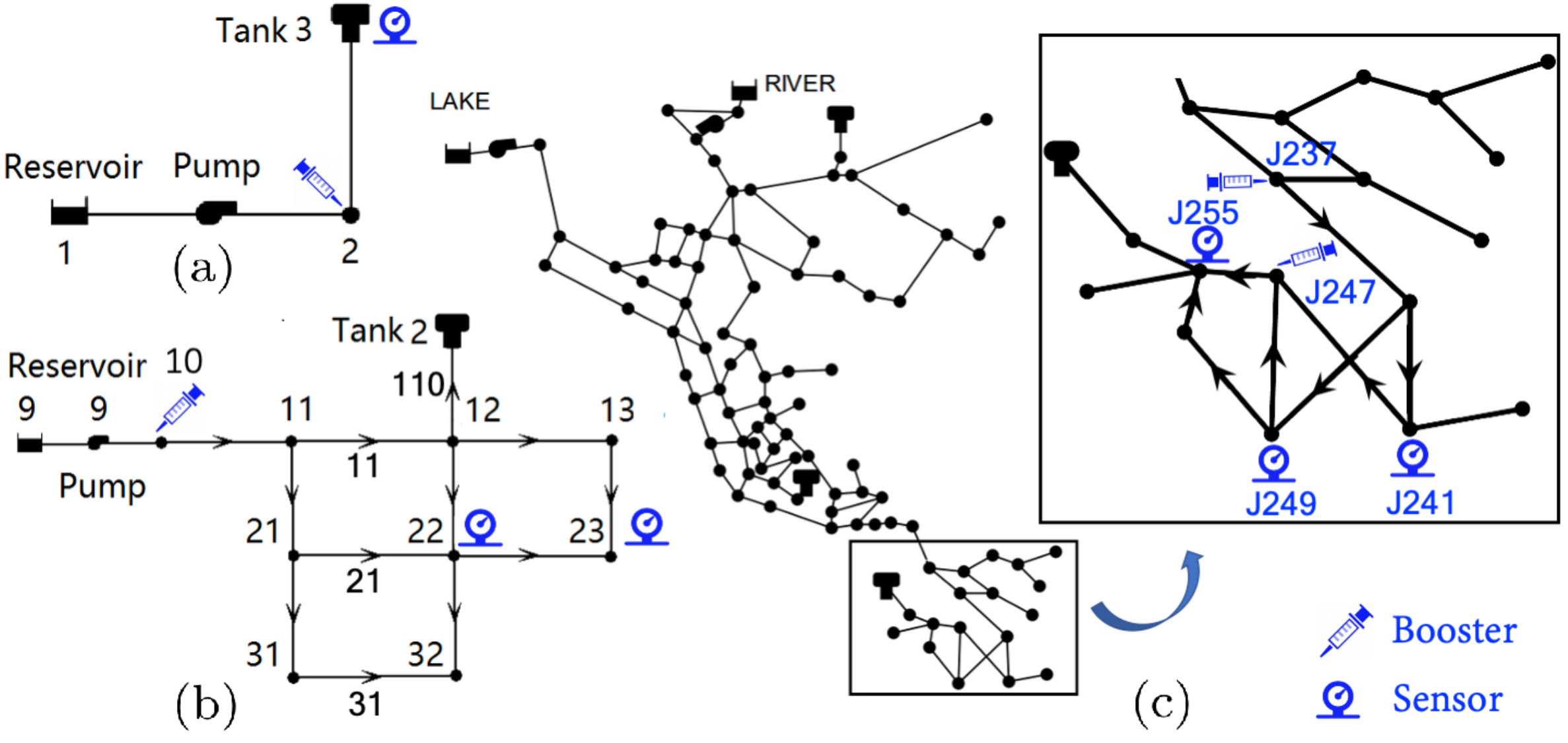}
	\caption{ Three tested networks: (a) three-node network; (b) Net1 network; (c) Net3 network (zoomed-in part is the controlled area).}
	\label{fig:networks}
\end{figure}
\subsection{Network settings, practical conditions, and validation}\label{sec:settings}

\subsubsection{Settings}
The basic information of the tested networks are in Tab.~\ref{tab:info}. The three-node network in Fig.~\ref{fig:networks}\textcolor{black}{a} is a self-designed network for illustrative purpose that includes one junction J2, one pipe P23, one pump PM12, one tank TK3, and one reservoir R1. A booster station and a chlorine concentration sensor are installed at J2 and TK3, which  indicates it is a single-input single-output (SISO) system ($n_u = n_y = 1$). The pipe P23 connecting J2 and TK3 is split into $s_{L} = 150$ segments according to L-W scheme in Fig.~\ref{fig:network-demand}a. Hence the dimension of the full-order system $n_x = 154$ ($n_{\mathrm{N}} = n_{\mathrm{J}}+n_{\mathrm{R}} + n_{\mathrm{TK}} = 3$, and $ n_{\mathrm{L}} = n_{\mathrm{P}}\times s_{L} +  n_{\mathrm{M}} = 151$). The dimensions and other parameters of the other networks are listed in Tab.~\ref{tab:info}.  

\subsubsection{Non-zero initial conditions}~\label{sec:non-zeros}
In a water quality simulation,  initial chlorine concentrations are typically non-zero (i.e., $\m x(0) \neq \m 0$). With that in mind, some MOR algorithms are designed with zero initial conditions.  To tailor the MOR algorithms for water quality dynamics with non-zero initial conditions, we adopt the method from~\cite{Heinkenschloss2011} as follows. Suppose we have a full-order model $(\m A, \m B, \m C, \m D)$ with $\m x_o = \m x(0) \neq \m 0$, let $\tilde{\m x} = \m x - \m x_o$, then 
	\begin{subequations}~\label{equ:augmentedfullmode}
		\begin{align}
		\tilde{\m x}(k+1) &= \tilde{\m A} \tilde{\m x}(k) + \tilde{\m B} \tilde{\m u}(k) ~\label{equ:augmentedfullmode_state}\\
		\tilde{\m y}(k) &= \tilde{\m C} \tilde{\m x}(k) + \tilde{\m D} \tilde{\m u}(k),
		\end{align}
\end{subequations} 
where $\tilde{\m A} = \m A$, $\tilde{\m B} = \begin{bmatrix}
\m B & \m A \m x_o
\end{bmatrix}$, $\tilde{\m C} = \m C$, $\tilde{\m D} = \begin{bmatrix}
\m D & \m C \m x_o
\end{bmatrix}$, and $\tilde{\m u} = \begin{bmatrix}
\m u  \\ 1
\end{bmatrix}$. That is, the non-zero initial condition is viewed as an input for the new augmented full-order model~\eqref{equ:augmentedfullmode} with $\tilde{\m x}_o = \m 0$ such that the introduced MOR algorithms can be applied directly without any modification. 

\subsubsection{Validation} To quantify the performance of MOR methods such as accuracy, the comparisons of step responses for the full-order model~\eqref{equ:fullmode}  and reduced-order model~\eqref{equ:reducedmode} are necessary. This is customary in MOR studies. The core idea of the designed experiments is applying the same input signal $\m u(k)$ to both the full- and reduced-order systems and then comparing output differences (i.e., step response errors between $\m y(k)$ and $\hat{\m y}(k) $). 
To quantify the accuracy of the MOR algorithms, we use  the root-mean-square error (RMSE) defined as
	\begin{align*}
	\mathrm{RMSE} = \sqrt{\frac{1}{m}\sum_{k=1}^{m}{|| \m y(k)- \hat{\m y}(k)||_2^2}}
	\end{align*}
to quantify the error between  full-order model output $\m y(k)$ and reduced-order model $\hat{\m y}(k)$ of a time horizon of $m$ time-steps. The RMSE is a well-known and understood metric to quantify errors between observations and predictions from a model.

\begin{figure}
	\centering
	\includegraphics[width=0.99\linewidth]{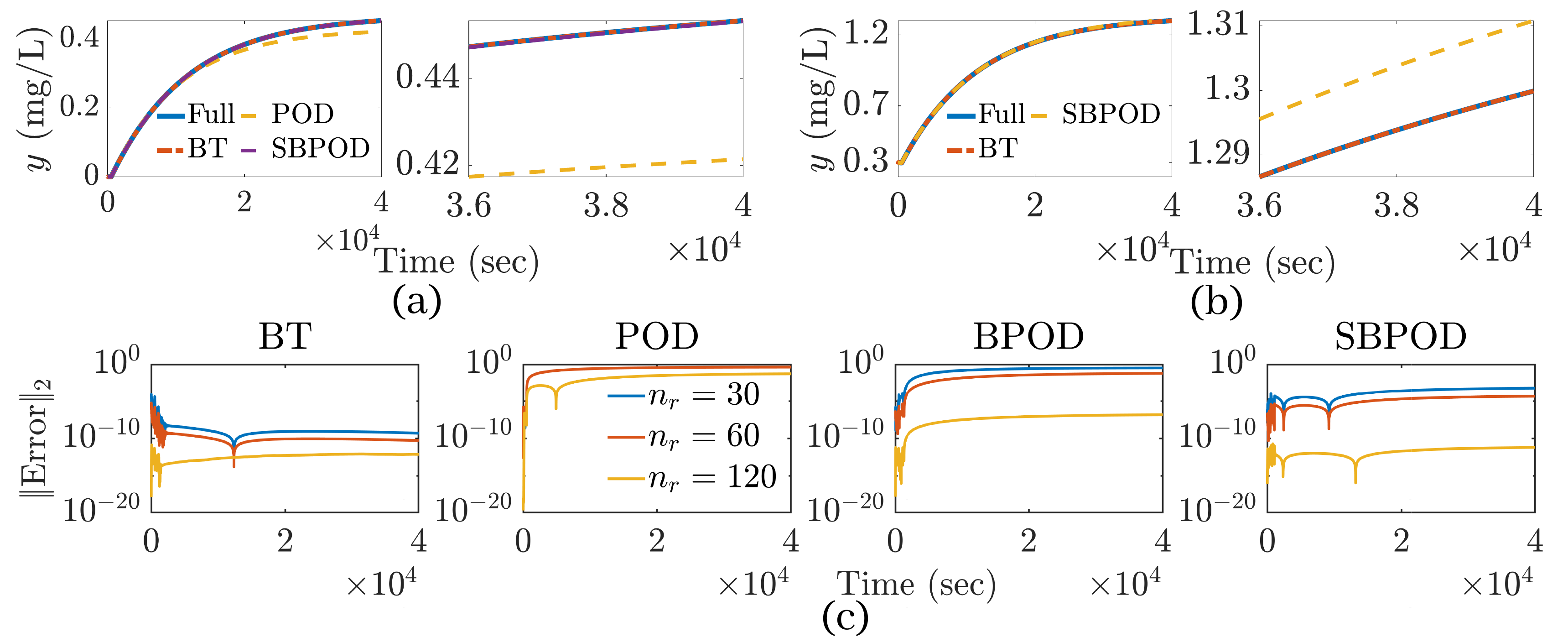}
	\caption{Step responses (with amplitude of 50) of full-/reduced- order models for the three-node network under: (a) zero-initial conditions; (b) non-zero initial conditions. (c) The step response errors between full-/reduced- order models under zero-initial conditions as $n_r$ increases from 30 to 120. Note that the adjacent figures in each of Figs.~\ref{fig:3-node}(a) and (b) are zoomed in versions, showcasing the differences in WQ MOR errors towards the end of the simulation. The log-scale of the errors is used to show differences in performance of the different algorithms. }
	\label{fig:3-node}
\end{figure}


\subsection{Accuracy and stability of MOR methods}
\subsubsection{The three-node network} The accuracy of MOR methods of the three-node network under zero initial conditions (i.e., $\m x_o = \m 0$) is presented in Fig.~\ref{fig:3-node}a(left). The input is a step signal with amplitude of 50, that is, we inject $50$ mg chlorine per sampling time ($\Delta t = 20$ seconds) at J2 in Fig.~\ref{fig:networks}\textcolor{black}{a}. The blue line  is the step response of the full-order model~\eqref{equ:fullmode} ($n_x = 154$), and the dotted lines are the step responses of the reduced-order models~\eqref{equ:reducedmode} obtained from BT, POD, and SBPOD ($n_r =30$, $118$, and $28$). The data is generated from Procedures~\ref{proc:BT},~\ref{proc:POD},~\ref{proc:BPOD}, and Algorithm~\ref{alg:generalMOR}. Fig.~\ref{fig:3-node}a(right) depicts the magnified step responses when time is limited  in $[36000, 40000]$ seconds. It can be seen that \textit{(i)} the reduced-order models from BT, POD, and SBPOD ($n_r =30$, $118$, and $28$) successfully maintain the input-output relationship with smaller orders ($n_r < n_x = 154$), and \textit{(ii)} at the end of the test ($[36000, 40000]$ sec), POD starts to show large errors while the results of BT and BPOD are still close to the full-order system. 

The accuracy test under non-zero initial conditions (i.e., $\m x_o \neq \m 0$) is also evaluated for the three-node network. All settings including the input signal remain the same except that the initial chlorine concentrations at  R1, J2, and TK3 are $1.0$ mg/L, $0.5$ mg/L, and $0.3$ mg/L; the initial chlorine concentration in PM12 and P23 are $0.75$ mg/L, and $0.3$ mg/L. The corresponding results are shown in Fig.~\ref{fig:3-node}b. Note that the output of POD is not shown due to it producing unreasonable results with huge $\mathrm{RMSE} = 23.51$. This is because the augmented system~\eqref{equ:augmentedfullmode} views non-zero initial conditions as inputs. In this way, the input matrix of~\eqref{equ:augmentedfullmode} $\tilde{\m B}$ includes two parts: the actual chlorine injection $\m B$ and \textit{virtual inputs} from the initial condition $\m A \m x_o$. If the majority of elements in vector $\m A \m x_o$ are non-zeros, then this system has too many inputs making the POD hard to capture the input-output relationship. This feature results in POD being unsuitable for for water quality dynamics with non-zero initial conditions. However, the non-zero initial conditions are not problematic for BT and BPOD since matrices $\m W_X$ or $\m H_m$ have the term $\m C \m B$ that cancels such effect. 



\begingroup
\setlength{\tabcolsep}{2pt} 
\renewcommand{\arraystretch}{1.3} 
\begin{table*}[]
	\centering
	\small
	\caption{Accuracy with (i) step signal as inputs, and (ii) non-zero (zero) initial conditions and corresponding computational time (in second) among three classical MOR methods for three tested networks. Different initial conditions are tested to showcase the performance of mainstream MOR algorithms compared to the proposed SBPOD. The computational time is also shown to illustrate the practical feasible of implementation.}
	\label{tab:Computational}
	\begin{tabular}{c|c|c|c|c|c|c}
		\hline
		\multirow{2}{*}{\textit{}} & \multicolumn{3}{c|}{\textit{\makecell{RMSEs  when applying step signals \\with $\m x_o= \m 0$ ($\m x_o \neq \m 0$)}}} & \multicolumn{3}{c}{\textit{\makecell{Computational time \\(in second)}}} \\ \cline{2-7} 
		& \textit{three-node network} & \textit{Net1} & \textit{Net3} & \textit{three-node network} & \textit{Net1} & \textit{Net3} \\ \hline
		\textit{BT} & \makecell{$1.75 \times 10^{-6}$ \\ ($1.71 \times 10^{-4}$)} &  \makecell{$1.63 \times 10^{-6}$  \\($6.7 \times 10^{-3}$)} & \makecell{NA$^*$ \\(NA$^*$)} & 0.245 & 17.2 & Impractical$^*$  \\ \hline
		\multicolumn{1}{l|}{\textit{POD}} & \makecell{$1.87 \times 10^{-2}$ \\(23.51)} & \makecell{$9.24 \times 10^{-5}$ \\($1.0 \times 10^{30}$)} & \makecell{$7.16 \times 10^{-5}$ \\(0.49)} & 0.219 & 6.4 & 245.8 \\ \hline
		\multicolumn{1}{l|}{\textit{SBPOD}} & \makecell{$1.54 \times 10^{-4}$ \\($4.2\times 10^{-3}$)} & \makecell{$6.9 \times 10^{-4}$ \\($6.5 \times 10^{-4}$)} & \makecell{$4.30 \times 10^{-6}$ \\($2.70 \times 10^{-3}$)} & 0.241 & 6.6 & 150.4 \\ \hline \hline
		\multicolumn{7}{l}{\footnotesize{
				\makecell{$^*$NA is ``Not Applicable" as BT fails to produce the reduced-order model after 96 hours of simulation.}}}
	\end{tabular}
\end{table*}
\endgroup

To further test the accuracy of all MOR methods (i.e., classical BT, POD, and BPOD and proposed SPOD) and the impact of parameter $n_r$, we show the  norm of step response errors (denoted by $\norm{\mathrm{Error}}_{2} = \norm{ \m y - \hat{\m y}}_{2}$) between full-order model~\eqref{equ:fullmode} and reduced-order models~\eqref{equ:reducedmode} under zero-initial conditions in Fig.~\ref{fig:3-node}c.
As $n_r$ increases from $30$ to $120$, the step response errors of all MOR methods decrease. We note the following. First, the amplitude of step response is in $10^{0}$ level (see Fig.~\ref{fig:3-node}c) while the error is in $10^{-5}$ or lower level for BT and SBPOD. This indicates that the error is tiny---it can be neglected and hence the reduced models are accurate. \textcolor{black}{Second, POD does not perform well even when $n_r$ reaches to $120$ ($n_x = 154$).} This is due to POD only considering the controllability Gramian in Section~\ref{sec:comparsion}. Third, classical BPOD (with $m = \underline{m}= 160$) performs worse than SBPOD ($m = 400 > \underline{m} = 160$); this indicates classical BPOD cannot ensure the stability or accuracy and is not suitable for water quality dynamics.  The value of parameter $\underline{m} = 160$ for this simple network can be obtained easily by Equation~\eqref{equ:travel}, that is, it equals the time-step from J2 to TK3. We show details of how to find $\underline{m}$ or $m$  using a more complex Net1 network in the next section.  All these test results corroborate the analyses performed in Section~\ref{sec:comparsion} and Tab.~\ref{tab:complexity}.

Furthermore, the $\mathrm{RMSEs}$ of all MOR methods (classical BT and POD, proposed SBPOD) for the three-node network under zero or non-zero initial conditions are shown in Tab.~\ref{tab:Computational}. The classical BPOD cannot ensure stability under either zero or non-zero initial conditions (see the further test in Section~\ref{sec:result-net1} for more details and discussions). Hence, the corresponding $\mathrm{RMSEs}$ are not included in the table. Based on these small $\mathrm{RMSEs}$, we conclude that MOR methods perform well under zero initial conditions while BT and SBPOD perform the best under the realistic non-zero initial conditions of water quality dynamics. 





\subsubsection{Net1 network}~\label{sec:result-net1} The accuracy tests of  classical BT, classical POD, and SBPOD methods for the Net1 network under zero/non-zero initial conditions are performed. Instead of presenting all overlapped lines in figures, we adopt $\mathrm{RMSEs}$ in Tab.~\ref{tab:Computational} to compare accuracy among all MOR methods. The full order of Net1 is $n_x = 1293$, and BT and SBPOD reduce the order to $n_r = 113$ and $n_r = 117$ under non-zero initial conditions.  Note that POD fails to provide accurate reduced-order models from the corresponding $\mathrm{RMSEs}$ shown in Tab.~\ref{tab:Computational}. For zero initial conditions, MOR methods (i.e., classical BT and POD and SBPOD) successfully reduce the system orders, and $n_r$ of BT, POD, and SBPOD are $145$, $270$, and $115$. From these small $\mathrm{RMSEs}$ in Tab.~\ref{tab:Computational}, it is evident that the reduced-order systems are accurate.

Next, we show how parameter $\underline{m}$ is obtained for Net1, and test the performance of BPOD (with an $m$ less than $\underline{m}$), SBPOD with the posterior stabilization method (same $m$ with BPOD), and SBPOD with the a priori stabilization method ($m$ larger than the $\underline{m}$).

\begin{figure*}
	\centering
	\includegraphics[width=0.75\linewidth]{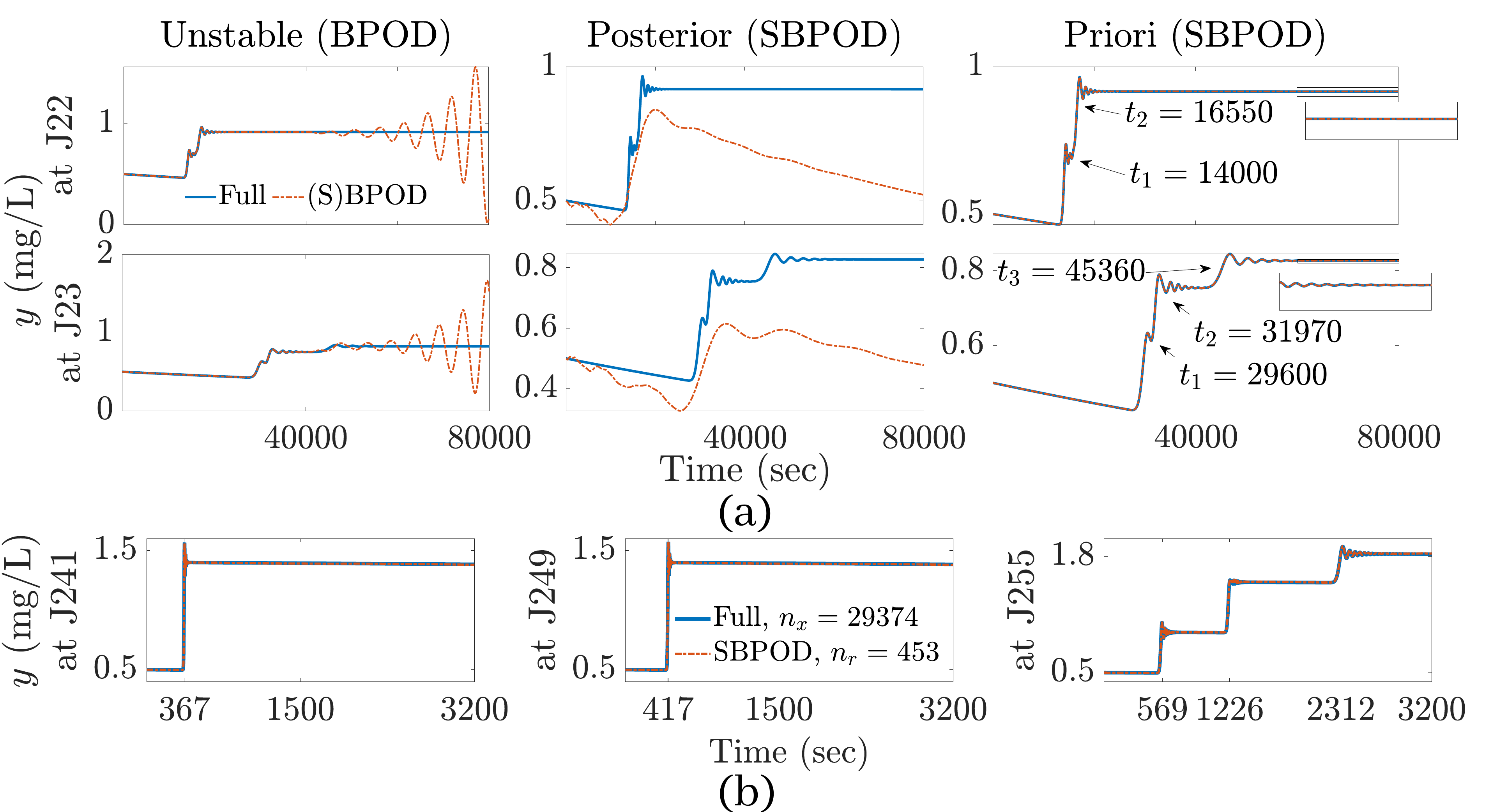}
	\caption{ (a) step responses from BPOD and SBPOD with two stabilization methods for Net1 under non-zero initial conditions; (b) step responses from different output locations of reduced-order models via BPOD methods for Net3 under  on-zero initial chlorine concentrations.}
	\label{fig:stablization}
\end{figure*}


%

\noindent \textit{\textbf{Computing lower bounds for} $m$.} The first step in generating stability-preserving reduced-order models is to compute the lower bound for $m$ for the SBPOD method. When the length of snapshot $m$ of BPOD is large enough for water quality dynamics, BPOD can preserve stability; otherwise, it cannot. This is the stability-preserving condition for BPOD as introduced in Section~\ref{sec:priori}. The two approaches to finding $\underline{m}$ are tested for the Net1 network.

The first method given in~\eqref{equ:travel} is used to obtain the proper $\underline{m}$ which requires computing the travel time of the last chlorine parcel injected at the booster J10 to the furthest sensor J23. The first (second) sensor installed at J22 (J23) receives two (three) step responses from the booster installed at J10, and the corresponding time to reach the peak is marked in the third column of Fig.~\ref{fig:stablization}a. The peak time of the last step response at J22 (J23) is $t_2 = 16550$ ($t_3 = 45360$), and the arrival or travel time of the last chlorine parcel (step response) is slightly less than the peak time. That is,  16000 for J22 and 39750 for J23 (we do not mark it in Fig.~\ref{fig:stablization}). Hence, parameter $\underline{m}$ can be obtained by the longest time 39750 divided by the sampling time $\Delta t = 15$ seconds. That is, parameter  $\underline{m}$ is set to $2650$ via the first method. 

We also choose Equation~\eqref{equ:condition}, that is the second method, to obtain the proper $\underline{m}$ which requires computing the dominant pole $p$. For Net1, $p = 0.9967 \pm 0.0563 i$, and $T_s = {-4}/{ln|p|}  = 2350.9$ resulting in $\underline{m}=2351$ for the second method. Note that we computed many decimals for the complex number $p$ and the natural logarithm of the magnitude $|p|$, resulting in $\underline{m}=2351$.

Although parameter $\underline{m}$ obtained by two methods are different, both are reasonable. After testing, we notice that when $m \in [2351, 2650]$, the reduced-order model for Net1 via BPOD shows instability as an oscillation tail is observed at the end of the step response; see the first column of figures in Fig.~\ref{fig:stablization}a. 

\noindent \textit{\textbf{Computing and simulating reduced-order models.}}  After finding the appropriate lower bounds for $m$, Algorithm~\ref{alg:generalMOR} is implemented to compute a stable reduced-order model for Net1. The step responses of the posterior stabilization method (stabilizing unstable reduced-order model from classical BPOD via solving an SDP~\eqref{equ:convex_stablization1}) are shown in the second column of Fig.~\ref{fig:stablization}a. Comparing with the unstable step responses from classical BPOD with $m = 2650$ (i.e., the first column of Fig.~\ref{fig:stablization}a), the yielding responses after posterior stabilization are not oscillatory but are unfortunately inaccurate. The step responses of the a priori method ($m$ is set to $4000$ that is larger than $2650$ found by stability-preserving condition) are shown in the third column of Fig.~\ref{fig:stablization}a. Compared with results of the posterior stabilization method,  the responses are stable and the method results in accurate estimation of the step response (see the zoom-in areas).



\subsubsection{Net3 network}  We present the results for a larger network, namely Net3 shown in Fig.~\ref{fig:networks}\textcolor{black}{c}, under zero ($\m x_o = 0$ mg/L) or non-zero initial chlorine concentrations ($\m x_o = 0.5$ mg/L).  Two booster locations (J237 and J247) in Fig.~\ref{fig:networks}\textcolor{black}{c} inject chlorine with a rate of $50$ mg and $40$ mg per sampling time ($\Delta t = 0.25$ second), that are two step signals with amplitudes of 50 and 40. Fig.~\ref{fig:stablization}b presents step responses of full-order model and reduced-model from SBPOD with non-zero initial conditions at three sensors (J255, J241, and J249). The full order of Net3 is $n_x = 29347$, and the reduced-order $n_r$ from  SBPOD is $453$. \textcolor{black}{Note that the BT method fails to produce a reduced-order model after 96 hours of simulation; thus, the results of BT are not included in Fig.~\ref{fig:stablization}b.} Furthermore, POD method is not suitable for non-zero initial condition case, and its results are not shown in Fig.~\ref{fig:stablization}b either. For zero initial condition case, both POD and SBPOD reduce system order successfully, and the reduced-orders $n_r$ are $689$ and $491$. Instead of presenting step response figures, we showcase the $\mathrm{RMSEs}$ of MOR methods in Tab.~\ref{tab:Computational}.


\subsection{Computational time}
Tab.~\ref{tab:Computational} presents the computational time for computing the reduced-order models for the three algorithms for different networks. We note that the computational time of classical BPOD and SBPOD are nearly identical (seeing that the computation of $m$ requires little to no time), and only results of SBPOD are shown for simplicity. The full-order $n_x$ and the length of snapshots $m$ for the three-node network are set as $154$ and $1000$ when testing computational time. For Net1, they are $1293$ and $7500$; for Net3, these values are $29374$ and $14000$.



\textcolor{black}{From Tab.~\ref{tab:Computational}, it can be seen that BT is impractical when $n_x$ is large since it fails to give a reduced-order model for Net3 after running the algorithm for 96 hours.} The computational time for SBPOD is less than the one by POD for Net3 which interestingly contradicts the theoretical complexity from Tab.~\ref{tab:complexity}. This happens because we implement a technique to avoid matrix-matrix multiplication by exploiting the structure of the block Hankel matrix in~\eqref{equ:svd}. Consequently, this results in a reduction of the computational time of the SBPOD method. The Github codes~\cite{wangMOR} also include a detailed description of how the problem structure is exploited to reduce the computational time. The details are omitted in this paper for brevity.

\subsection{SBPOD-Based MPC vs. Full-Order MPC}
One of the main objectives of investigating MOR algorithms for water quality dynamics, besides the ability to perform forward state-simulations more efficiently, is to showcase the algorithms' performance in the context of real-time feedback control (i.e., controlling chlorine concentrations $\m x(k)$ through the installed booster station controllers $\m u(k)$) or state observation and estimation (i.e., estimating the states $\m x(k)$ via noisy sensor measurements $\m y(k)$). To that end, the objective of this section is to test the applicability of the presented SBPOD-generated, reduced-order model within a model predictive control (MPC) framework that regulates chlorine concentrations through controlling dosages of injected chlorine at the booster stations. We specifically focus on SBPOD herein as it produces the best performance (in terms of accuracy under different initial conditions, computational time, and stability-preserving) among the other tested methods. 

The motivation for using the reduced-order model is as follows: the regulation of chlorine concentrations with strict state constraints (e.g., as the ones mandated by the US EPA) in a water network can be posed as a constrained MPC optimization---as we have shown in our recent work~\cite{wang2020effective}. We have also showcased that the control routine for the full-order system becomes computationally intractable when considering MPC with state- and input-constraints and a large control horizon (e.g., 30 minutes) even for a mid-size network (e.g, Net1). To solve this issue, we have relaxed the strict constraints and converted the intractable constrained MPC into an unconstrained and tractable one. However, this leads to another issue. That is, the unconstrained MPC cannot guarantee the state and input bounds, which is one of the limitations of our previous work~\cite{wang2020effective}. In this paper, we investigate the following research question in this section:

\begin{quote}
	\textit{Can the water system operator utilize the SBPOD-generated, reduced-order model with constrained MPC to determine the desired dosages of injected chlorine? How would such an application compare to using a similar control routine for the full-order model? Would utilizing the reduced-order model result in a significant increase in the operation cost of controlling chlorine concentrations? How does that impact the controlled chlorine consecrations?   } 
\end{quote} 

To answer these questions, we use the Net1 network as an illustration. Similar performance is exhibited for other networks. Note that we set a small control horizon parameter, that is 5 minutes, for Net1 to make optimization problems with full-order models tractable so that the control performance of full-/reduced- order models based constrained MPC can be tested and compared. In particular, we consider the WQ MPC to be constrained by lower and upper
bounds constraints on chlorine concentrations in all links and nodes as follows: $c_{\min} = 0.2 \, \mathrm{mg/L}$ and $c_{\max}=4\,  \mathrm{mg/L}$, that is
$\m x(t) \in [\m c_{\min} ,\; \m c_{\max} ]$. These constraints are common for WQ regulation and are set by the US EPA. Then, the MPC from~\cite{wang2020effective} is solved for both the reduced order model as well as the full-order one and subsequently compared in terms of performance and computational time.  First, the full-order models are computed for a time horizon of 24 hours. This is then followed by computing the SBPOD-generated reduced-order models via the a priori stabilization method. Second, the optimal control signals of chlorine dosages for both the full- and reduced-order models are computed via the constrained MPC from study~\cite{wang2020effective}. The reference signal for the MPC algorithm to track is set as 1.5 mg/L. That is, the goal of the controller is to maintain the chlorine concentration at around 1.5 mg/L at J22 and J23 (the sensor nodes in Net1) by controlling the injected mass rate of chlorine $\m u(k)$ at J10 (the only control node in Net1). Finally, the control signals are then applied to the water quality simulation in the Net1 network. For brevity, we do not reproduce the details here; the interested reader can examine the provided Github codes for the parameters and settings~\cite{wangMOR}. 

\begin{figure}
	\centering
	\includegraphics[width=0.95\linewidth]{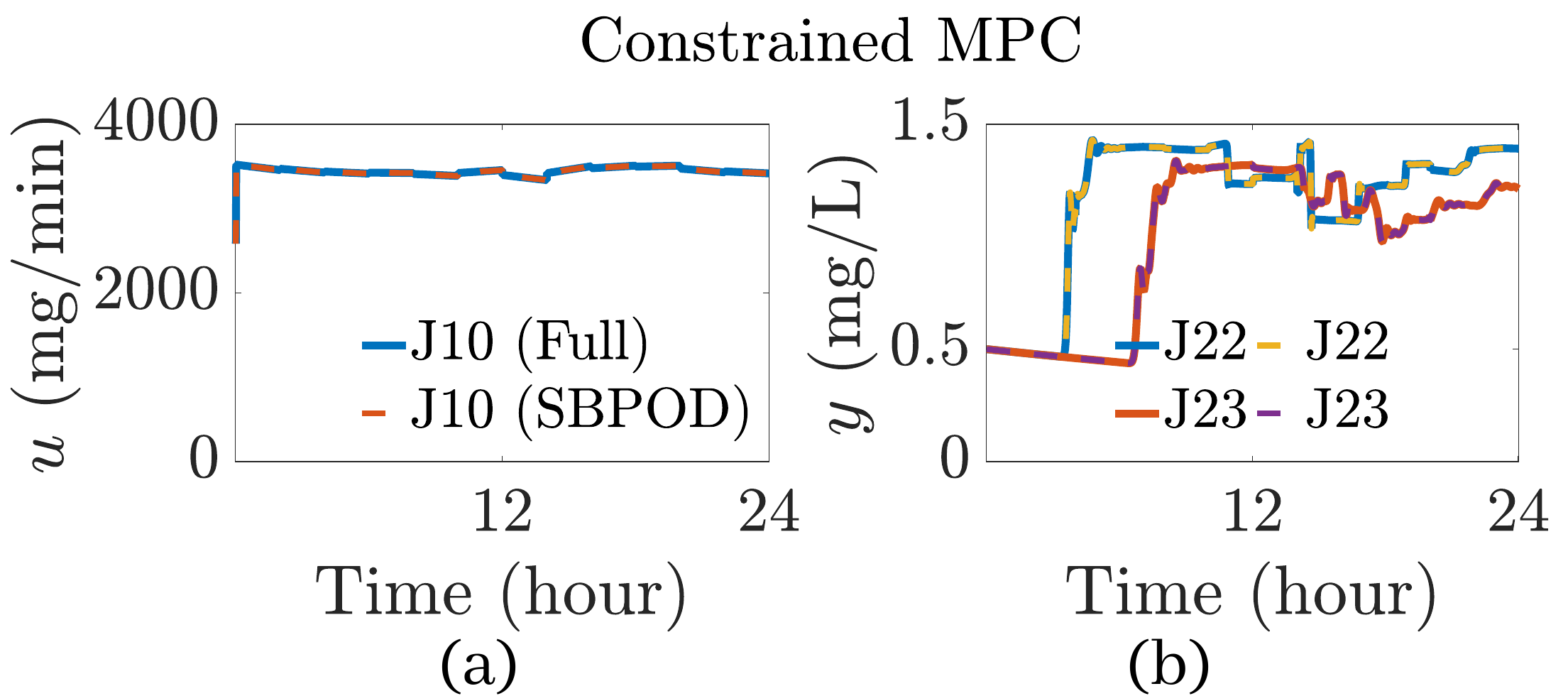}
	\caption{The comparisons of  (a) control actions and (b) subsequent chlorine concentrations solved by the full- and reduced-order model (i.e., SBPOD) for the constrained MPC problems. We note here that the chlorine dosage schedules are time-varying, and these are computed via the two MPC formulations for the specific booster station in this case study.}
	\label{fig:mpcbpodnet1}
\end{figure}
%

To assess the performance of the SBPOD-based chlorine dosage constrained model predictive controller, in comparison with the similar MPC that utilizes the more computationally expensive full-order one, we test for three metrics: the computational time of solving optimization problems, the difference in the cost function of running the controllers (i.e., the operational cost of running the controller), and the evolution of the chlorine concentration. The three metrics are assessed for the two approaches: \textit{(i)} the SBPOD-based constrained MPC  and \textit{(ii)} the standard constrained MPC for the full-order model. Fig.~\ref{fig:mpcbpodnet1}a and Fig.~\ref{fig:mpcbpodnet1}b show the injected mass rate $\m u(t)$  and the corresponding chlorine concentrations at J22 and J23 for \textit{(i)} and \textit{(ii)} for the entire time horizon for both the reduced- and full-order models. 
The $\mathrm{RMSEs}$ for control input $\m u$ and output $\m y$ between the full-order model and reduced-order one are $5.95 \times 10^{-2}$ and $3.63 \times 10^{-6}$. The computational time of solving constrained MPC problems with full- and reduced-order models are  $50.7$ seconds and $31.7$ seconds. Both operational costs from these two different controller are $4.49 \times 10^5$.

By examining the results, we observe that the reduced-order model produces nearly identical performance (in terms of chlorine concentration dynamics and the optimal chlorine dosages) when compared with the full-order one, while incurring smaller computational time. 
\textcolor{black}{We note here that solving constrained MPC with full-order models for Net3, a larger network than Net1, is computationally \textcolor{black}{impractical}. After a running time of 12 hours that are clearly larger than the prediction horizon, solutions are still not obtained since this optimization problem has at least $n_x = 29374$ optimization variables and millions of constraints.} However, with the reduced-order model, the constrained optimization problem only has $n_r = 453$ optimization variables and at around ten thousand constraints making it computationally tractable. In short, the number of optimization variables and constraints of constrained MPC with reduced-order models is small regardless of the size of the network/full-order model since parameter $n_r$ is usually within a thousand after MOR. The reduced-order model enables water system operators to use constrained MPC that guarantees the input and output bounds while remaining tractable for larger networks.

\section{Paper Summary, Limitations and Future Work}~\label{sec:limitations}

The presented research in this paper explores the potential of model order reduction for water quality dynamics in drinking water networks. The presented stability-preserving BPOD (SBPOD) algorithm is more usable than its counterparts in the literature of MOR due to its accuracy, low computational burden, tolerance to initial conditions, and stability-insuring property. Compared with the classical BT method, SBPOD is computationally tractable while yielding accurate estimates of output measurements of chlorine concentrations. Furthermore, compared to POD or BPOD, SBPOD handles a variety of initial conditions while ensuring stability and accuracy of estimated outputs from the reduced-order models. Finally, we demonstrate that when using the proposed SBPOD within an MPC framework to control chlorine concentrations, the water system operator can use the SBPOD-generated reduced-order model instead of the full-order one without incurring any significant losses in the operational cost of controlling the network, while saving orders of magnitude in computational time. In particular, the provided Github codes can be scaled through a third party to guide the operator in using MOR-based control for water quality regulation. In particular, this makes control algorithms such as MPC amenable to real-time implementation---as a result of utilizing the reduced-order models.  In short, and through the comprehensive testing and theoretical discussions, interested water quality experts can utilize the SBPOD to perform accurate simulations and constrained, feedback control for water quality dynamics. 

The presented research in this paper comes with its limitations which are outlined next jointly with the authors' future research directions. First, in-depth and large-scale simulations for WQ MOR reduction in large-scale water networks---with thousands of pipes and junctions and under various testing conditions, hydraulic profiles and contamination events---is needed. The objective of the present paper is to explore the literature's state-of-the-art and identify its weaknesses, and propose a simple approach to mitigate the weaknesses. Hence, a significant effort is needed to understand how the proposed approach perform when more realistic networks are considered. The provided Github codes~\cite{wangMOR} can help the interested readers in further testing and calibrating our algorithms. 

 Second, the proposed two methods to obtain a stable reduced-order model have some disadvantages. For example, the a posteriori method still needs to solve an optimization problem while the accuracy of a stabilized reduced-order model is not guaranteed.  Although this optimization problem is a convex one with a size of the reduced order model system, investigating its solutions for larger systems would be an interesting research direction.   Third, the length of snapshots in the a priori method for a generic case is based on finding the dominant poles of the full-order model via a known empirical formula. To the authors' best knowledge, this formula has not been verified with thorough theoretical analysis. Further analysis is needed. 

 Finally, the paper only studies MOR for linear water quality models of single species and as a result, multi-species models that are often nonlinear is outside the scope of the paper. Future research directions include investigating model order reduction algorithms for water quality dynamics that involve multi-species interactions, in comparison with models of focusing on single-species ones. A state-space model for water quality dynamics in a multi-species framework would likely yield a nonlinear state-space description that then necessitate  designing MOR algorithms for nonlinear dynamics. To that end, relevant future work will focus on \textit{(i)} thorough modeling of multi-species reaction dynamics and \textit{(ii)} reducing such dynamics to lower order models.



\section*{Acknowledgments}
This material is based upon work supported by the National Science Foundation under Grants 1728629, 2015603, 2015671, 2151392, and 2015658. The authors also acknowledge the comments and suggestions offered by the editor and the two reviewers. All the codes and tested networks are freely available in open data repository~\cite{wangMOR} for research reproducibility.

\bibliographystyle{IEEEtran}
\bibliography{mybib}

\end{document}

%% file: preambleT.tex
\usepackage{epsfig,color,amsmath}
\usepackage[noadjust]{cite}
\usepackage{amsthm}
\IEEEoverridecommandlockouts
\usepackage{wrapfig} 
\usepackage{bm}
\usepackage{epstopdf}
\usepackage{makecell}
\usepackage{amssymb}
\usepackage{lipsum}  
\usepackage{url}
\usepackage{enumitem}
\usepackage{multirow}
\usepackage{makecell}
\setlist[itemize]{leftmargin=*}

\DeclareMathOperator{\subjectto}{subject\ to}

\makeatother

\newcommand{\m}{\boldsymbol}
\allowdisplaybreaks[4]
\newcommand{\mbb}[1]{\mathbb{#1}}

\usepackage{graphicx}
\usepackage{ifpdf}
\usepackage{epstopdf}
\usepackage{ifpdf}
\ifpdf
\DeclareGraphicsExtensions{.eps}
\else
\DeclareGraphicsExtensions{.eps}
\fi
\usepackage[caption=off]{subfig} 

\usepackage[colorlinks = true,
linkcolor = blue,
urlcolor  = blue,
citecolor = blue,
anchorcolor = blue]{hyperref}
\usepackage{stackengine} 
\stackMath
\usepackage[flushleft]{threeparttable}
\usepackage{makecell}

\usepackage{multicol}
\usepackage{mathtools}

\usepackage[normalem]{ulem}
\usepackage{mathrsfs}
\DeclareMathAlphabet\mathbfcal{OMS}{cmsy}{b}{n}
\DeclarePairedDelimiter\abs{\lvert}{\rvert}%
\DeclarePairedDelimiter\norm{\lVert}{\rVert}%

\usepackage{stackengine}

\usepackage{etoolbox}
\usepackage[procnumbered,linesnumbered,lined,boxed,commentsnumbered,ruled,longend]{algorithm2e}


\makeatletter
\AtBeginEnvironment{procedure}{\let\c@algocf\c@procedure}
\makeatother

%% file: MOR-arxiv.bbl
\begin{thebibliography}{10}
\providecommand{\url}[1]{#1}
\csname url@samestyle\endcsname
\providecommand{\newblock}{\relax}
\providecommand{\bibinfo}[2]{#2}
\providecommand{\BIBentrySTDinterwordspacing}{\spaceskip=0pt\relax}
\providecommand{\BIBentryALTinterwordstretchfactor}{4}
\providecommand{\BIBentryALTinterwordspacing}{\spaceskip=\fontdimen2\font plus
\BIBentryALTinterwordstretchfactor\fontdimen3\font minus
  \fontdimen4\font\relax}
\providecommand{\BIBforeignlanguage}[2]{{%
\expandafter\ifx\csname l@#1\endcsname\relax
\typeout{** WARNING: IEEEtran.bst: No hyphenation pattern has been}%
\typeout{** loaded for the language `#1'. Using the pattern for}%
\typeout{** the default language instead.}%
\else
\language=\csname l@#1\endcsname
\fi
#2}}
\providecommand{\BIBdecl}{\relax}
\BIBdecl

\bibitem{wang2020effective}
S.~Wang, A.~F. Taha, and A.~A. Abokifa, ``How effective is model predictive
  control in real-time water quality regulation? state-space modeling and
  scalable control,'' \emph{Water Resources Research}, 2020, in press.

\bibitem{boulos2006comprehensive}
P.~Boulos, K.~Lansey, and B.~Karney, \emph{Comprehensive Water Distribution
  Systems Analysis Handbook for Engineers and Planners}.\hskip 1em plus 0.5em
  minus 0.4em\relax MWH Soft, Incorporated, 2006.

\bibitem{rowley2005model}
C.~W. Rowley, ``Model reduction for fluids, using balanced proper orthogonal
  decomposition,'' \emph{International Journal of Bifurcation and Chaos},
  vol.~15, no.~03, pp. 997--1013, 2005.

\bibitem{Baur2014}
U.~Baur, P.~Benner, and L.~Feng, ``{Model Order Reduction for Linear and
  Nonlinear Systems: A System-Theoretic Perspective},'' \emph{Archives of
  Computational Methods in Engineering}, vol.~21, no.~4, pp. 331--358, 2014.

\bibitem{Willcox2002}
K.~Willcox and J.~Peraire, ``{Balanced model reduction via the proper
  orthogonal decomposition},'' \emph{AIAA Journal}, vol.~40, no.~11, pp.
  2323--2330, 2002.

\bibitem{Montier2017}
L.~Montier, T.~Henneron, B.~Goursaud, and S.~Clenet, ``{Balanced Proper
  Orthogonal Decomposition Applied to Magnetoquasi-Static Problems Through a
  Stabilization Methodology},'' \emph{IEEE Transactions on Magnetics}, vol.~53,
  no.~7, 2017.

\bibitem{adamjan1971analytic}
V.~M. Adamjan, D.~Z. Arov, and M.~Kre{\u\i}n, ``{Analytic properties of Schmidt
  pairs for a Hankel operator and the generalized Schur-Takagi problem},''
  \emph{Mathematics of the USSR-Sbornik}, vol.~15, no.~1, p.~31, 1971.

\bibitem{Glover1984}
K.~Glover, ``All optimal hankel-norm approximations of linear multivariable
  systems and their {$L^\infty$} error bounds,'' \emph{International journal of
  control}, vol.~39, no.~6, pp. 1115--1193, 1984.

\bibitem{moore1981principal}
B.~Moore, ``Principal component analysis in linear systems: Controllability,
  observability, and model reduction,'' \emph{IEEE transactions on automatic
  control}, vol.~26, no.~1, pp. 17--32, 1981.

\bibitem{sirovich1987turbulence}
L.~Sirovich, ``Turbulence and the dynamics of coherent structures. i. coherent
  structures,'' \emph{Quarterly of applied mathematics}, vol.~45, no.~3, pp.
  561--571, 1987.

\bibitem{grimme1997krylov}
E.~Grimme, ``Krylov projection methods for model reduction,'' Ph.D.
  dissertation, 1997.

\bibitem{antoulas1990solution}
A.~Antoulas, J.~Ball, J.~Kang, and J.~Willems, ``On the solution of the minimal
  rational interpolation problem,'' \emph{Linear Algebra and its Applications},
  vol. 137, pp. 511--573, 1990.

\bibitem{beattie2008interpolation}
C.~A. Beattie and S.~Gugercin, ``Interpolation theory for structure-preserving
  model reduction,'' in \emph{2008 47th IEEE Conference on Decision and
  Control}.\hskip 1em plus 0.5em minus 0.4em\relax IEEE, 2008, pp. 4204--4208.

\bibitem{Bai2002}
Z.~Bai, ``{Krylov subspace techniques for reduced-order modeling of large-scale
  dynamical systems},'' \emph{Applied Numerical Mathematics}, vol.~43, no. 1-2,
  pp. 9--44, 2002.

\bibitem{gallivan2006model}
K.~Gallivan, A.~Vandendorpe, and P.~Van~Dooren, ``Model reduction and the
  solution of sylvester equations,'' \emph{MTNS, Kyoto}, vol.~50, 2006.

\bibitem{gugercin2008iterative}
S.~Gugercin, ``An iterative svd-krylov based method for model reduction of
  large-scale dynamical systems,'' \emph{Linear Algebra and its Applications},
  vol. 428, no. 8-9, pp. 1964--1986, 2008.

\bibitem{lall2003error}
S.~Lall and C.~Beck, ``Error-bounds for balanced model-reduction of linear
  time-varying systems,'' \emph{IEEE Transactions on Automatic Control},
  vol.~48, no.~6, pp. 946--956, 2003.

\bibitem{lumley2007stochastic}
J.~L. Lumley, \emph{Stochastic tools in turbulence}.\hskip 1em plus 0.5em minus
  0.4em\relax Courier Corporation, 2007.

\bibitem{astrid2008missing}
P.~Astrid, S.~Weiland, K.~Willcox, and T.~Backx, ``Missing point estimation in
  models described by proper orthogonal decomposition,'' \emph{IEEE
  Transactions on Automatic Control}, vol.~53, no.~10, pp. 2237--2251, 2008.

\bibitem{ulanicki1996simplification}
B.~Ulanicki, A.~Zehnpfund, and F.~Martinez, ``Simplification of water
  distribution network models,'' in \emph{Proc., 2nd Int. Conf. on
  Hydroinformatics}.\hskip 1em plus 0.5em minus 0.4em\relax Balkema Rotterdam,
  Netherlands, 1996, pp. 493--500.

\bibitem{shamir2008optimal}
U.~Shamir and E.~Salomons, ``Optimal real-time operation of urban water
  distribution systems using reduced models,'' \emph{Journal of Water Resources
  Planning and Management}, vol. 134, no.~2, pp. 181--185, 2008.

\bibitem{preis2011efficient}
A.~Preis, A.~J. Whittle, A.~Ostfeld, and L.~Perelman, ``Efficient hydraulic
  state estimation technique using reduced models of urban water networks,''
  \emph{Journal of Water Resources Planning and Management}, vol. 137, no.~4,
  pp. 343--351, 2011.

\bibitem{perelman2008water}
L.~Perelman and A.~Ostfeld, ``Water distribution system aggregation for water
  quality analysis,'' \emph{Journal of water resources planning and
  management}, vol. 134, no.~3, pp. 303--309, 2008.

\bibitem{lax1964difference}
P.~D. Lax and B.~Wendroff, ``Difference schemes for hyperbolic equations with
  high order of accuracy,'' \emph{Communications on pure and applied
  mathematics}, vol.~17, no.~3, pp. 381--398, 1964.

\bibitem{chen2013linear}
\BIBentryALTinterwordspacing
C.~Chen, \emph{Linear System Theory and Design}, ser. The Oxford Series in
  Electrical and Computer Engineering.\hskip 1em plus 0.5em minus 0.4em\relax
  Oxford University Press, Incorporated, 2013. [Online]. Available:
  \url{https://books.google.com/books?id=XyPAoAEACAAJ}
\BIBentrySTDinterwordspacing

\bibitem{Himpe2018}
C.~Himpe, ``{emgr-The empirical Gramian framework},'' \emph{Algorithms},
  vol.~11, no.~7, pp. 1--27, 2018.

\bibitem{Himpe2020}
\BIBentryALTinterwordspacing
------, ``{Comparing (Empirical-Gramian-Based) Model Order Reduction
  Algorithms},'' 2020. [Online]. Available:
  \url{http://arxiv.org/abs/2002.12226}
\BIBentrySTDinterwordspacing

\bibitem{lall1999empirical}
S.~Lall, J.~E. Marsden, and S.~Glava{\v{s}}ki, ``Empirical model reduction of
  controlled nonlinear systems,'' \emph{IFAC Proceedings Volumes}, vol.~32,
  no.~2, pp. 2598--2603, 1999.

\bibitem{horn2012matrix}
R.~A. Horn and C.~R. Johnson, \emph{Matrix analysis}.\hskip 1em plus 0.5em
  minus 0.4em\relax Cambridge university press, 2012.

\bibitem{amsallem2012stabilization}
D.~Amsallem and C.~Farhat, ``Stabilization of projection-based reduced-order
  models,'' \emph{International Journal for Numerical Methods in Engineering},
  vol.~91, no.~4, pp. 358--377, 2012.

\bibitem{li2002low}
J.-R. Li and J.~White, ``Low rank solution of lyapunov equations,'' \emph{SIAM
  Journal on Matrix Analysis and Applications}, vol.~24, no.~1, pp. 260--280,
  2002.

\bibitem{pan1999complexity}
V.~Y. Pan and Z.~Q. Chen, ``The complexity of the matrix eigenproblem,'' in
  \emph{Proceedings of the thirty-first annual ACM symposium on Theory of
  computing}, 1999, pp. 507--516.

\bibitem{feng2018faster}
X.~Feng, W.~Yu, and Y.~Li, ``Faster matrix completion using randomized svd,''
  in \emph{2018 IEEE 30th International Conference on Tools with Artificial
  Intelligence (ICTAI)}.\hskip 1em plus 0.5em minus 0.4em\relax IEEE, 2018, pp.
  608--615.

\bibitem{prajna2003pod}
S.~Prajna, ``Pod model reduction with stability guarantee,'' in \emph{42nd IEEE
  International Conference on Decision and Control (IEEE Cat. No. 03CH37475)},
  vol.~5.\hskip 1em plus 0.5em minus 0.4em\relax IEEE, 2003, pp. 5254--5258.

\bibitem{Rowley2004}
C.~W. Rowley, T.~Colonius, and R.~M. Murray, ``{Model reduction for
  compressible flows using POD and Galerkin projection},'' \emph{Physica D:
  Nonlinear Phenomena}, 2004.

\bibitem{Barone2009}
M.~F. Barone, I.~Kalashnikova, D.~J. Segalman, and H.~K. Thornquist, ``{Stable
  Galerkin reduced order models for linearized compressible flow},''
  \emph{Journal of Computational Physics}, 2009.

\bibitem{kalashnikova2010stability}
I.~Kalashnikova and M.~Barone, ``On the stability and convergence of a galerkin
  reduced order model (rom) of compressible flow with solid wall and far-field
  boundary treatment,'' \emph{International journal for numerical methods in
  engineering}, vol.~83, no.~10, pp. 1345--1375, 2010.

\bibitem{kalashnikova2012stable}
I.~Kalashnikova and S.~Arunajatesan, ``A stable galerkin reduced order model
  for compressible flow,'' in \emph{10th World Congress on Computational
  Mechanics (WCCM), WCCM-2012-19407, Sao Paulo, Brazil}, 2012.

\bibitem{sirisup2004spectral}
S.~Sirisup and G.~E. Karniadakis, ``A spectral viscosity method for correcting
  the long-term behavior of pod models,'' \emph{Journal of Computational
  Physics}, vol. 194, no.~1, pp. 92--116, 2004.

\bibitem{wang2012proper}
Z.~Wang, I.~Akhtar, J.~Borggaard, and T.~Iliescu, ``Proper orthogonal
  decomposition closure models for turbulent flows: a numerical comparison,''
  \emph{Computer Methods in Applied Mechanics and Engineering}, vol. 237, pp.
  10--26, 2012.

\bibitem{kalashnikova2014stabilization}
I.~Kalashnikova, B.~van Bloemen~Waanders, S.~Arunajatesan, and M.~Barone,
  ``Stabilization of projection-based reduced order models for linear
  time-invariant systems via optimization-based eigenvalue reassignment,''
  \emph{Computer Methods in Applied Mechanics and Engineering}, vol. 272, pp.
  251--270, 2014.

\bibitem{Bond2008}
B.~N. Bond and L.~Daniel, ``{Guaranteed stable projection-based model reduction
  for indefinite and unstable linear systems},'' in \emph{IEEE/ACM
  International Conference on Computer-Aided Design, Digest of Technical
  Papers, ICCAD}, 2008.

\bibitem{benosman2017robust}
M.~Benosman, J.~Borggaard, and B.~Kramer, ``Robust pod model stabilization for
  the 3d boussinesq equations based on lyapunov theory and extremum seeking,''
  in \emph{2017 American Control Conference (ACC)}.\hskip 1em plus 0.5em minus
  0.4em\relax IEEE, 2017, pp. 1827--1832.

\bibitem{lyapunov1992general}
A.~M. Lyapunov, ``The general problem of the stability of motion,''
  \emph{International journal of control}, vol.~55, no.~3, pp. 531--534, 1992.

\bibitem{rossman2000epanet}
L.~A. Rossman \emph{et~al.}, ``{EPANET} 2: users manual,'' 2000.

\bibitem{Eliades2016}
D.~G. Eliades, M.~Kyriakou, S.~Vrachimis, and M.~M. Polycarpou, ``Epanet-matlab
  toolkit: An open-source software for interfacing epanet with matlab,'' in
  \emph{Proc. 14th International Conference on Computing and Control for the
  Water Industry (CCWI)}, The Netherlands, Nov 2016, p.~8.

\bibitem{wangMOR}
\BIBentryALTinterwordspacing
ShenWang9202, ``{ShenWang9202/MOR4WQM: MOR FOR Water quality dynamics},'' Feb.
  2021. [Online]. Available: \url{https://doi.org/10.5281/zenodo.4554174}
\BIBentrySTDinterwordspacing

\bibitem{Heinkenschloss2011}
\BIBentryALTinterwordspacing
M.~Heinkenschloss, T.~Reis, and A.~C. Antoulas, ``{Balanced truncation model
  reduction for systems with inhomogeneous initial conditions},''
  \emph{Automatica}, vol.~47, no.~3, pp. 559--564, 2011. [Online]. Available:
  \url{http://dx.doi.org/10.1016/j.automatica.2010.12.002}
\BIBentrySTDinterwordspacing

\end{thebibliography}
